\documentclass[aps,12pt,prd,tightenlines,floatfix,showpacs,superscriptaddress]{revtex4-1}

\usepackage[dvips]{graphicx}
\usepackage[english]{babel}
\usepackage{amsmath}
\usepackage{amssymb}
\usepackage{slashed}
\allowdisplaybreaks

\begin{document}

\title{Gauge Invariant Linear Response Theory of Relativistic BCS superfluids}

\author{Hao Guo}
\affiliation{Department of Physics, Southeast University, Nanjing 211189, China.}
\affiliation{Department of Physics, University of Hong Kong, Hong Kong, China.}
\email{guohao.ph@gmail.com}
\author{Chih-Chun Chien}
\affiliation{Theoretical Division, Los Alamos National Laboratory, Los Alamos, NM, 87545, U.S.A.}
\email{chihchun@lanl.gov}
\author{Yan He}
\affiliation{Department of Physics, University of California, Riverside, CA 92521, U.S.A.}
\email{heyan@ucr.edu}
\date{\today}


\begin{abstract}
We develop a gauge-invariant linear response theory for relativistic Bardeen-Cooper-Schrieffer (BCS) superfluids based on a consistent-fluctuation-of-the order-parameter (CFOP) approach. The response functions from the CFOP approach satisfy important generalized Ward identities. The gauge invariance of the CFOP theory is a consequence of treating the gauge transformation and the fluctuations of the order parameter on equal footing so collective-mode effects are properly included. We demonstrate that the pole of the response functions is associated with the massless Goldstone boson. Important physical quantities such as the compressibility and superfluid density of relativistic BCS superfluids can also be inferred from our approach. We argue that the contribution from the massless Goldstone boson is crucial in obtaining a consistent expression for the compressibility.
\end{abstract}

\maketitle

\section{Introduction}
The Bardeen-Cooper-Schrieffer theory of fermionic superfluids (see \cite{Schrieffer_book} for a review on its application to conventional superconductors) has provided not only an explanation for conventional superconductors and other fermionic superfluids but also a paradigm for studying macroscopic quantum coherence due to interactions. Soon after its discovery, the challenge of how to cast its formalism in a gauge invariant form when a charged system is interacting with an electromagnetic (EM) field has drawn broad interest. At the linear-response-theory level, there have been two major approaches. Nambu in his seminal paper \cite{Nambu60} formulated this challenge in terms of generalized Ward identities (GWIs) and proposed an approach based on a set of integral equations for finding a gauge-invariant EM vertex which governs the kernels of response functions. This approach has been reviewed in Schrieffer's book on superconductivity \cite{Schrieffer_book} and also finds applications in other branches of physics such as nuclear matter interacting with neutrinos \cite{ReddyPRC04}.

There is another approach based on consistent fluctuations of the order parameter (CFOP), which is the main theme of this paper. In this approach, the effects of gauge transformation from the gauge field are balanced by the fluctuations of the order parameter in a consistent fashion. This is made possible by treating the terms induced by the gauge field as well as the fluctuations of the order parameter equally in the perturbative Hamiltonian. Although the kernels of response functions from this approach are not a solution of Nambu's integral equations, the CFOP formalism is manifestly gauge invariant and we will show that GWIs are satisfied. This is because being a solution of Nambu's integral equations is only a sufficient condition for satisfying the GWIs, but not a necessary condition. Importantly, this approach reproduces the compressibility correctly as that from the equations of state and this demonstrates self-consistency of the CFOP approach.

The theory of CFOP has an interesting history. Kadanoff and Martin \cite{Kadanoff61} first proposed this approach in a less complete form by considering only the phase fluctuations of the order parameter. Their idea is to decompose the three-particle Green's function in a way that can respect gauge invariance. Betbeder Matibet and Nozieres \cite{Nozieres69} and Kulik et al. \cite{KulikJLTP81} independently formulated this approach in more complete forms with both phase and amplitude fluctuations later on. This approach has also been formulated by the Keldysh formalism with time-ordered Green's functions in Ref.~\cite{Arseev}. After its successful application to conventional superconductors,  a generalization of this approach to nonrelativistic fermionic superfluids satisfies important sum rules and has been applied to ultra-cold atomic systems \cite{HaoPRL10,HaoPRL11,HaoNJP11}. It has also been discussed in the physics of neutron stars \cite{Gusakov10}. Here we base on our formalism of a relativistic version of the BCS theory \cite{OurNPA09} with Kulik's approach to CFOP and develop a gauge-invariant linear response theory of relativistic fermionic BCS superfluids. To demonstrate the versatility of this approach, we will address the collective mode associated with the massless Goldstone boson in the symmetry-broken phase, the density susceptibility which governs the compressibility, and the superfluid density.

To further contrast these two approaches, we also present the relativistic version of Nambu's integral equations for the EM vertex. There have been attempts to find an iterated solution based on the random phase approximation (RPA) \cite{ReddyPRC04}. In nonrelativistic BCS superfluids it is possible to argue that the RPA-based theory satisfies the corresponding GWI. To our knowledge, neither a relativistic version of Nambu's integral equations nor a complete proof of the gauge invariance of the RPA-based linear response theory for relativistic BCS superfluids have been explicitly presented. Since a major goal of this paper is to advocate the CFOP theory of relativistic Fermi superfluids, we will limit our discussions on Nambu's integral-equation approach.

This paper is organized as the following. Sec.~\ref{chap:1} briefly reviews a microscopic theory for relativistic BCS superfluids that will be the foundation of this work. Sec.~\ref{chap:3} presents the CFOP formalism and we explain in more detail how our theory respects gauge invariance in Sec.~\ref{chap:5}. Sec.~\ref{chap:6} gives the explicit expressions of the response functions from our CFOP approach. Sec.~\ref{chap:coll}, Sec.~\ref{chap:dndmu}, and Sec.~\ref{chap:Meissner} show some applications of the CFOP theory to the collective modes, compressibility, and the Meissner effect for a relativistic BCS superfluid. We briefly discuss a relativistic version of Nambu's integral-equation approach and its associated GWIs in Sec.~\ref{chap:Nambu}. Sec.~\ref{chap:conclusion} concludes our work. Some details and conventions are given in the Appendix.

\section{Microscopic Theory of Relativistic Fermi Superfluids}\label{chap:1}
Several relativistic models of a two-component BCS superfluids have been formulated in Refs.~\cite{NishidaPRD05,OhsakuPRB01,ZhuangPRD07a,AbukiNPA05,WangPRD07,OurNPA09} and we briefly review the model following the BCS-Leggett mean field theory \cite{ZhuangPRD07a,OurNPA09} without any external gauge field here. The Lagrangian density is
\begin{equation} \label{eqn:L}
\mathcal{L}(\mathbf{x})=\sum_{\sigma=\uparrow,\downarrow}\bar{\psi}_{\sigma}(i\gamma^{\mu}\partial_{\mu}-m+\mu\gamma^{0})\psi_{\sigma}+\mathcal{L}_{I}(\mathbf{x}),
\end{equation}
where $\psi$, $\bar{\psi}$ are Dirac spinors which denote the fermion fields with mass $m$ and chemical potential $\mu$. The representation of the $\gamma-$matrix and some useful properties are given in Appendix~\ref{app:spinor}. Throughout this paper, we take the convention $e=c=\hbar=1$ and use $\sigma$ to denote the pseudo-spin $\uparrow$ and $\downarrow$ with $\uparrow=-\downarrow$ and $\bar{\sigma}=-\sigma$. The pseudo-spin may refer to some internal degrees of freedom such as the color indices in quantum chromodynamics. $\mathcal{L}_{I}$ describes the attractive pairing interactions between particles with different pseudo-spins and it takes the form \cite{NishidaPRD05}
\begin{equation}
\mathcal{L}_{I}(\mathbf{x})=g(\psi^{T}_{\uparrow}Ci\gamma_{5}\psi_{\downarrow})(\bar{\psi}_{\downarrow}i\gamma_{5}C\bar{\psi}^{T}_{\uparrow}),
\end{equation}
where $g$ is the attractive coupling constant, and the charge conjugation matrix $C$ is defined as $C=i\gamma_{0}\gamma_{2}$. The gap function that is also the order parameter is given by $\Delta(\mathbf{x})=g\langle\psi^{T}_{\uparrow}Ci\gamma_{5}\psi_{\downarrow}\rangle$. The standard BCS approximation then gives
\begin{eqnarray}
\mathcal{L}_{BCS}(\mathbf{x})=\sum_{\sigma=\uparrow,\downarrow}\bar{\psi}_{\sigma}(i\gamma^{\mu}\partial_{\mu}-m+\mu\gamma^{0})\psi_{\sigma}+\Delta^{*}(\psi^{T}_{\uparrow}Ci\gamma_{5}\psi_{\downarrow})+\Delta(\bar{\psi}_{\downarrow}i\gamma_{5}C\bar{\psi}^{T}_{\uparrow}).
\end{eqnarray}
The corresponding form of the Hamiltonian density is then given by
\begin{eqnarray} \label{eqn:H}
\mathcal{H}_{BCS}(\mathbf{x})=\sum_{\sigma=\uparrow,\downarrow}\bar{\psi}_{\sigma}(-i\vec{\gamma}\cdot\nabla+m-\mu\gamma^{0})\psi_{\sigma}-\Delta^{*}(\psi^{T}_{\uparrow}Ci\gamma_{5}\psi_{\downarrow})-\Delta(\bar{\psi}_{\downarrow}i\gamma_{5}C\bar{\psi}^{T}_{\uparrow}).
\end{eqnarray}
In the broken-symmetry phase the order parameter may be chosen to be real. Here we present our theory in Matsubara formalism, which is applicable to both zero and finite temperature $T$. We will focus on $T=0$ results and a generalization to finite $T$ within the BCS approximation is straightforward.
 To simplify the notation, we group the imaginary time $\tau=i t$ and $\mathbf{x}$ as a four-vector $x=(\tau,\mathbf{x})$ and define
\begin{eqnarray}\label{eqn:HO}
\mathcal{O}(x)=e^{H_{BCS}\tau}\mathcal{O}(\mathbf{x})e^{-H_{BCS}\tau}
\end{eqnarray}
where $H_{BCS}=\int d^3\mathbf{x}\mathcal{H}_{BCS}(\mathbf{x})$. The single particle Green's function and anomalous Green's function are given by
\begin{eqnarray}
G(x,x')=-\langle T_{\tau}[\psi_{\uparrow}(x)\bar{\psi}_{\uparrow}(x')]\rangle, \mbox{  } F(x,x')=-\langle T_{\tau}[\psi_{\uparrow}(x)\psi^T_{\downarrow}(x')C]\rangle,
\end{eqnarray}
where $T_{\tau}$ denotes the $\tau$-order of operators. When $H_{BCS}$ is time-independent, $G$ and $F$ depend only on the difference $\tau-\tau'$. Let $x^+=(\tau+0^+,\mathbf{x})$. The gap function can be expressed as
\begin{eqnarray}\label{E8}
\Delta(\mathbf{x})=g\textrm{Tr}\big[i\gamma_5F(x,x^+)\big].
\end{eqnarray}
The number density for each species is defined by $n_{\sigma}(\mathbf{x})=\langle\bar{\psi}_{\sigma}(\mathbf{x})\gamma^0\psi_{\sigma}(\mathbf{x})\rangle=\langle\psi_{\sigma}^{\dagger}(\mathbf{x})\psi_{\sigma}(\mathbf{x})\rangle$. It can be also calculated from the single particle Green's function $n_{\sigma}(\mathbf{x})=\textrm{Tr}\big[\gamma^0G(x,x^+)\big]$. Therefore, the total fermion number is given by
\begin{eqnarray}\label{eqn:n3}
n=\int d^3\mathbf{x}(n_{\uparrow}(\mathbf{x})+n_{\downarrow}(\mathbf{x}))=2\int d^3\mathbf{x}\textrm{Tr}\big[\gamma^0G(x,x^+)\big].
\end{eqnarray}
It can be shown that the Green's function and anomalous Green's function satisfy the following equations of motion
\begin{eqnarray}
& &[-\gamma^{0}\partial_{\tau}+i\vec{\gamma}\cdot\nabla-(m-\mu\gamma^{0})]G(x,x^{\prime})+i\Delta\gamma_{5}\widetilde{F}(x,x^{\prime})=\delta(x-x^{\prime})\mathbf{1}_{4\times4},\\ & &[-\gamma^{0}\partial_{\tau}+i\vec{\gamma}\cdot\nabla-(m-\mu\gamma^{0})]F(x,x^{\prime})+i\Delta\gamma_{5}\widetilde{G}(x,x^{\prime})=0,
\end{eqnarray}
where $\widetilde{G}(x,x^{\prime})=CG^T(x',x)C$, $\widetilde{F}(x,x^{\prime})=\gamma^0F^{\dagger}(x',x)\gamma^0$, $\mathbf{1}_{4\times4}$ is the four-dimensional identity matrix, and $\delta(x-x')=\delta(\tau-\tau')\delta(\mathbf{x}-\mathbf{x}')$. Here we define the fermion four-momentum at finite temperature as $P=(i\omega_n,\mathbf{p})$, where $\omega_n$ is the fermionic Matsubara frequency given by $\omega_n=(2n+1)\pi k_{B}T$, where $k_{B}$ is the Boltzmann constant. The quasi-particle energies are given by $E^{\pm}_{\mathbf{p}}=\sqrt{\xi^{\pm2}_{\mathbf{p}}+\Delta^2}$ with   $\xi^{\pm}_{\mathbf{p}}=\epsilon_{\mathbf{p}}\pm\mu$ and $\epsilon_{\mathbf{p}}=\sqrt{\mathbf{p}^2+m^2}$. With the help of the energy projectors
\begin{equation}
\Lambda_{\pm}(\mathbf{p})=\frac{1}{2}[1\pm\frac{\gamma^{0}(\vec{\gamma}\cdot\mathbf{p}+m)}{\epsilon_{\mathbf{p}}}],
\end{equation}
the solution of $G$ and $F$ in momentum space are
\begin{equation}\label{eqn:G}
G(P,\mu)=\big[\frac{u^{-2}_{\mathbf{p}}\Lambda_{+}(\mathbf{p})}{i\omega_{n}-E^{-}_{\mathbf{p}}}
+\frac{v^{-2}_{\mathbf{p}}\Lambda_{+}(\mathbf{p})}{i\omega_{n}+E^{-}_{\mathbf{p}}}
+\frac{u^{+2}_{\mathbf{p}}\Lambda_{-}(\mathbf{p})}{i\omega_{n}+E^{+}_{\mathbf{p}}}
+\frac{v^{+2}_{\mathbf{p}}\Lambda_{-}(\mathbf{p})}{i\omega_{n}-E^{+}_{\mathbf{p}}}\big]\gamma^{0},
\end{equation}
\begin{equation}\label{eqn:F}
F(P,\mu)=\big[\frac{u^{-}_{\mathbf{p}}v^{-}_{\mathbf{p}}\Lambda_{+}(\mathbf{p})}{i\omega_{n}-E^{-}_{\mathbf{p}}}
-\frac{u^{-}_{\mathbf{p}}v^{-}_{\mathbf{p}}\Lambda_{+}(\mathbf{p})}{i\omega_{n}+E^{-}_{\mathbf{p}}}
+\frac{u^{+}_{\mathbf{p}}v^{+}_{\mathbf{p}}\Lambda_{-}(\mathbf{p})}{i\omega_{n}-E^{+}_{\mathbf{p}}}
-\frac{u^{+}_{\mathbf{p}}v^{+}_{\mathbf{p}}\Lambda_{-}(\mathbf{p})}{i\omega_{n}+E^{+}_{\mathbf{p}}}\big]i\gamma_{5},
\end{equation}
where $u^{\pm 2}_{\mathbf{p}}=\frac{1}{2}(1+\frac{\xi^{\pm}_{\mathbf{p}}}{E^{\pm}_{\mathbf{p}}})$ and $v^{\pm 2}_{\mathbf{p}}=\frac{1}{2}(1-\frac{\xi^{\pm}_{\mathbf{p}}}{E^{\pm}_{\mathbf{p}}})$. The energy projectors $\Lambda_{+}(\mathbf{p})$ and $\Lambda_{-}(\mathbf{p})$ project out the contributions from the fermion and anti-fermion respectively. In the nonrelativistic limit where $|\mathbf{p}|\ll m$, $|\mu-m|\ll m$ and $\Delta\ll m$, one gets $\Lambda_{+}(\mathbf{p})\simeq1$ and $\Lambda_{-}(\mathbf{p})\simeq0$ so our expressions reduce to the well-known nonrelativistic results. By taking the Fourier transform of Eq.(\ref{eqn:n3}), the fermion number is given by $n=2\sum_{P}\mbox{Tr}[\gamma^{0}G(P,\mu)]$, which is the number difference between the fermions and anti-fermions
\begin{equation}\label{ENUM}
n=n_+-n_-=4\sum_{\mathbf{p}}\big[u^{-2}_{\mathbf{p}}f(E^{-}_{\mathbf{p}})+v^{-2}_{\mathbf{p}}f(-E^{-}_{\mathbf{p}})\big]
-4\sum_{\mathbf{p}}\big[(u^{+2}_{\mathbf{p}}f(E^{+}_{\mathbf{p}})+v^{+2}_{\mathbf{p}}f(-E^{+}_{\mathbf{p}}))\big],
\end{equation}
where $n_{\pm}$ denote the density of fermion and anti-fermion. The Fermi momentum $k_F$ is defined by $n=2k^3_F/(3\pi^2)$, and the Fermi energy is $\epsilon_F=\sqrt{k^2_F+m^2}$. The Fourier transform of Eq.~(\ref{E8}) gives $\Delta=g\sum_{P}\mbox{Tr}[i\gamma_5F(P,\mu)]$, which leads to the gap equation
\begin{equation}\label{EGAP}
\frac{1}{g}=\sum_{\mathbf{p}}\big(\frac{1-2f(E^{-}_{\mathbf{p}})}{E^{-}_{\mathbf{p}}}
+\frac{1-2f(E^{+}_{\mathbf{p}})}{E^{+}_{\mathbf{p}}}\big).
\end{equation}
In the nonrelativistic limit, the number equation (\ref{ENUM}) and gap equation (\ref{EGAP}) reduce to the well-known nonrelativistic results except a factor of $2$ on both right-hand sides. This factor of $2$ comes from the fact that we have introduced the pseudo-spin, which brings two times more degrees of freedom.

Since the model is not renormalizable in $3+1$ dimensions, a regularization or a momentum cutoff $\Lambda$ is needed.
The relativistic limit of the BCS state depends on the Compton wavelength $\lambda_c=m^{-1}$ \cite{ZhuangPRD07a}. If $k_F\gg 1/\lambda_c=m$, the system evolves into the relativistic regime. It has been shown that this model can be generalized to describe the BCS - Bose-Einstein condensation (BEC)- relativistic BEC crossover \cite{ZhuangPRD07a} of Fermi gases.

To further compactify our expressions, we reformulate the relativistic BCS theory in the Nambu formalism \cite{Nambu60,Schrieffer_book}. This is more convenient for the discussions on the linear response to an external electromagnetic field. We introduce the Nambu-Gorkov spinors
\begin{displaymath}
\Psi(\mathbf{x})=
\begin{pmatrix}\displaystyle
\psi_{\uparrow}(\mathbf{x})\\
C\bar{\psi}^{T}_{\downarrow}(\mathbf{x})
\end{pmatrix}
,\mbox{ }\bar{\Psi}(\mathbf{x})=(\bar{\psi}_{\uparrow}(\mathbf{x}),\mbox{ }\psi^{T}_{\downarrow}(\mathbf{x})C).
\end{displaymath}
Moreover, we define
\begin{eqnarray}
\sigma_+=\frac{1}{2}(\sigma_1+i\sigma_2),\sigma_-=\frac{1}{2}(\sigma_1-i\sigma_2),\bar{\sigma}_+=\frac{1}{2}(\sigma_0+\sigma_3), \bar{\sigma}_-=\frac{1}{2}(\sigma_0-\sigma_3).
\end{eqnarray}
in Nambu space. One can show that $\Delta=g\langle\bar{\Psi}i\gamma_5\sigma_-\Psi\rangle$ and its Fourier transform is $\Delta_{\mathbf{q}}=g\sum_{\mathbf{p}}\langle\bar{\Psi}_{\mathbf{p}}i\gamma_5\sigma_-\Psi_{\mathbf{p}+\mathbf{q}}\rangle$. Similarly one can show that
\begin{eqnarray}\label{LMF}
\mathcal{L}_{BCS}=\bar{\Psi}\big(i\gamma^{\mu}\partial_{\mu}-m+\mu\gamma_0\sigma_3\big)\Psi+\bar{\Psi}\big(\Delta i\gamma_5\sigma_++\Delta^*i\gamma_5\sigma_-\big)\Psi
\end{eqnarray}
The Lagrangian density can be written as $\mathcal{L}_{BCS}=\bar{\Psi}\hat{G}^{-1}\Psi$, where the inverse propagator in momentum space is given by
\begin{eqnarray}\label{eqn:FP}
\hat{G}^{-1}(P,\mu)&=&-\langle T_{\tau}[\Psi_{\mathbf{p}}\bar{\Psi}_{\mathbf{p}}]\rangle=(i\omega_n+\mu\sigma_3)\gamma^0-\vec{\gamma}\cdot\mathbf{p}-m+\Delta i\gamma_5\sigma_1.
\end{eqnarray}

After evaluating the inverse of the right-hand-side, one gets the expression of the propagator (see Appendix \ref{appB})
\begin{eqnarray} \label{eqn:BG}
& &\hat{G}(P,\mu)=\nonumber\\
&=&\Big[\big(\frac{u^{-2}_{\mathbf{p}}}{i\omega_n-E^-_{\mathbf{p}}}+\frac{v^{-2}_{\mathbf{p}}}{i\omega_n+E^-_{\mathbf{p}}}\big)\Lambda_+(\mathbf{p})
+\big(\frac{u^{+2}_{\mathbf{p}}}{i\omega_n+E^+_{\mathbf{p}}}+\frac{v^{+2}_{\mathbf{p}}}{i\omega_n-E^+_{\mathbf{p}}}\big)\Lambda_-(\mathbf{p})\Big]\gamma^0\bar{\sigma}_+ \nonumber\\
&+&\Big[\big(\frac{u^{+2}_{\mathbf{p}}}{i\omega_n-E^+_{\mathbf{p}}}+\frac{v^{+2}_{\mathbf{p}}}{i\omega_n+E^+_{\mathbf{p}}}\big)\Lambda_+(\mathbf{p})
+\big(\frac{u^{-2}_{\mathbf{p}}}{i\omega_n+E^-_{\mathbf{p}}}+\frac{v^{-2}_{\mathbf{p}}}{i\omega_n-E^-_{\mathbf{p}}}\big)\Lambda_-(\mathbf{p})\Big]\gamma^0\bar{\sigma}_- \nonumber\\
&+&\Big[\frac{\Lambda_+(\mathbf{p})\Delta }{(i\omega_n)^2-E^{-2}_{\mathbf{p}}}
+\frac{\Lambda_-(\mathbf{p})\Delta }{(i\omega_n)^2-E^{+2}_{\mathbf{p}}}\Big]i\gamma_5\sigma_+
+\Big[\frac{\Lambda_+(\mathbf{p})\Delta }{(i\omega_n)^2-E^{+2}_{\mathbf{p}}}
+\frac{\Lambda_-(\mathbf{p})\Delta }{(i\omega_n)^2-E^{-2}_{\mathbf{p}}}\Big]i\gamma_5\sigma_-.
\end{eqnarray}
From Eqs.~(\ref{eqn:G}) and (\ref{eqn:F}), one finds that
\begin{eqnarray}\label{eqn:BG2}
& &\hat{G}(P,\mu)=\left( \begin{array}{ccc} G(P,\mu) & F(P,\mu) \\ F(P,-\mu) & G(P,-\mu) \end{array} \right).
\end{eqnarray}
Moreover, the number equation (\ref{ENUM}) and gap equation (\ref{E8}) can also be rewritten in Nambu space as
\begin{eqnarray}\label{eqn:n2}
n=\sum_P\mbox{Tr}\big[\sigma_3\gamma^0\hat{G}(P,\mu)\big],\quad\Delta=\frac{g}{2}\sum_P\mbox{Tr}\big[\sigma_1i\gamma_5\hat{G}(P,\mu)\big].
\end{eqnarray}
The expression (\ref{eqn:BG}) of the propagator in Nambu space can be further simplified to a more instructive form. We define the operator $\hat{E}_{\mathbf{p}}=\gamma^0(\vec{\gamma}\cdot\mathbf{p}+m)-\mu\sigma_3-\Delta\gamma^0i\gamma_5\sigma_1$ in Nambu space and also introduce the projectors
\begin{eqnarray}
\hat{\Lambda}_+(\mathbf{p})\equiv\left[ \begin{array}{ccc} \Lambda_+(\mathbf{p}) & 0 \\ 0 & \Lambda_-(\mathbf{p})\end{array}\right], \mbox{  } \hat{\Lambda}_-(\mathbf{p})\equiv1-\hat{\Lambda}_+(\mathbf{p})=\left[ \begin{array}{ccc} \Lambda_-(\mathbf{p}) & 0 \\ 0 & \Lambda_+(\mathbf{p})\end{array}\right].
\end{eqnarray}
Then the propagator becomes (see Appendix.\ref{app:C})
\begin{eqnarray}\label{eqn:BG3}
\hat{G}(P,\mu)=\big[\frac{\hat{u}^{-2}_{\mathbf{p}}}{i\omega_n-E^-_{\mathbf{p}}}+\frac{\hat{v}^{-2}_{\mathbf{p}}}{i\omega_n+E^-_{\mathbf{p}}}
+\frac{\hat{u}^{+2}_{\mathbf{p}}}{i\omega_n+E^+_{\mathbf{p}}}+\frac{\hat{v}^{+2}_{\mathbf{p}}}{i\omega_n-E^+_{\mathbf{p}}}\big]\gamma^0,
\end{eqnarray}
where the coefficients are given by
\begin{eqnarray}
\hat{u}^{\pm}_{\mathbf{p}}=\frac{(E^{\pm}_{\mathbf{p}}\mp\hat{E}_{\mathbf{p}})\hat{\Lambda}_{\mp}(\mathbf{p})}{2E^{\pm}_{\mathbf{p}}},
\hat{v}^{\pm}_{\mathbf{p}}=\frac{(E^{\pm}_{\mathbf{p}}\pm\hat{E}_{\mathbf{p}})\hat{\Lambda}_{\mp}(\mathbf{p})}{2E^{\pm}_{\mathbf{p}}}.
\end{eqnarray}
Those coefficients are the counterparts (in Nambu space) of the coefficients $u^{\pm}_{\mathbf{p}}$ and $v^{\pm}_{\mathbf{p}}$.

\section{Gauge-Invariant Linear Response Theory}\label{chap:3}
We consider fermions and anti-fermions coupled to a weak external EM field $A_{\mu}(\mathbf{x})$. The derivative $\partial_{\mu}$ in the Lagrangian density (\ref{eqn:L}) should be replaced by the covariant derivative $D_{\mu}=\partial_{\mu}+iA_{\mu}(\mathbf{x})$, which results in the interacting term $\mathcal{L}_{A}(\mathbf{x})=-\sum_{\sigma=\uparrow,\downarrow}\bar{\psi}_{\sigma}\gamma^{\mu}\psi_{\sigma}A_{\mu}=-J^{\mu}A_{\mu}$. In Nambu space, one can show that $\mathcal{L}_A=-\bar{\Psi}\gamma^{\mu}\sigma_3\Psi A_{\mu}$.

The corresponding Hamiltonian density is $\mathcal{H}_{A}(\mathbf{x})=-\mathcal{L}_{A}(\mathbf{x})$. Gauge invariance of a microscopic linear response theory with respect to an external EM field is made possible by considering the perturbations due to the fluctuations of the order parameter in a consistent fashion. The nonrelativistic version of the CFOP method has been extensively studied \cite{KulikJLTP81,KosztinPRB00,ZhaPRB95} and here we will develop a relativistic version of this method.

In equilibrium, the order parameter is given by $\Delta$. We assume that the deviation of the order parameter from its equilibrium is small and denote the small perturbation by $\Delta'(x)$. Therefore, $\Delta$ in Eq.(\ref{LMF}) is replaced by $\Delta\rightarrow\Delta+\Delta'$. Then the Hamiltonian density splits into two parts: the equilibrium expression and the part containing the deviation. Explicitly, $\mathcal{H}_{BCS}=\mathcal{H}_{BCS0}+\mathcal{H}'$
where
\begin{eqnarray}
\mathcal{H}_{BCS0}=\bar{\Psi}\big(-i\vec{\gamma}\cdot\nabla+m-\mu\gamma^{0}\sigma_3-\Delta i\gamma_5\sigma_1\big)\Psi
\end{eqnarray}
and
\begin{eqnarray}\label{eqn:H'}
\mathcal{H}'=\bar{\Psi}\big(\Delta_1i\gamma_5\sigma_1+\Delta_2i\gamma_5\sigma_2+\slashed{A}\sigma_3\big)\Psi.
\end{eqnarray}
Here $\Delta'=-(\Delta_1-i\Delta_2)$ and $\Delta'^*=-(\Delta_1+i\Delta_2)$. $\Delta_1$ and $\Delta_2$ are the negative real and imaginary parts of the fluctuations of the order parameter. The Hamiltonian $H_{BCS}$ becomes
\begin{eqnarray}
& &H_{BCS}=H_{BCS0}+H'  \\
&=&\sum_{\mathbf{p}}\bar{\Psi}_{\mathbf{p}}\big(\vec{\gamma}\cdot\mathbf{p}+m-\mu\gamma^{0}\sigma_3-\Delta i\gamma_5\sigma_1\big)\Psi_{\mathbf{p}}
+\sum_{\mathbf{p}\mathbf{q}}\bar{\Psi}_{\mathbf{p}+\mathbf{q}}\big(\Delta_{1\mathbf{q}}i\gamma_5\sigma_1+\Delta_{2\mathbf{q}}i\gamma_5\sigma_2+\slashed{A}_{\mathbf{q}}\sigma_3\big)\Psi_{\mathbf{p}}\nonumber
\end{eqnarray}
The interaction term may be considered as a scalar product $\hat{\mathbf{\Phi}}^T_{\mathbf{q}}\cdot\hat{\mathbf{\Sigma}}$, where
\begin{eqnarray}\label{GGP}
\hat{\mathbf{\Phi}}_{\mathbf{q}}=(\Delta_{1\mathbf{q}},\Delta_{2\mathbf{q}},A_{\mu\mathbf{q}})^T,\mbox{  } \hat{\mathbf{\Sigma}}=(\sigma_1i\gamma_5,\sigma_2i\gamma_5,\sigma_3\gamma^{\mu})^T,
\end{eqnarray}
are the generalized external potential and the generalized vertex function. To calculate the linear response of a relativistic Fermi superfluid to the perturbation $H'$, we introduce the response-function vector $\vec{\eta}$:
\begin{eqnarray}
\vec{\eta}(\tau,\mathbf{q})=\sum_{\mathbf{p}}\langle\bar{\Psi}_{\mathbf{p}}(\tau)\hat{\mathbf{\Sigma}}\Psi_{\mathbf{p}+\mathbf{q}}(\tau)\rangle,
\end{eqnarray}
where $\eta^{\mu}_3\equiv J^{\mu}$ corresponds to the current due to the external field and $\eta_{1,2}$ denote the perturbations due to the fluctuations of the gap function. The covariant index $\mu$ should not be confused with the chemical potential. The linear response theory is then written in a matrix form
\begin{eqnarray}\label{eqn:Q}
\vec{\eta}(\tau,\mathbf{q})&=&\tensor{Q}(\tau,\mathbf{q})\cdot\hat{\mathbf{\Phi}}_{\mathbf{q}} \nonumber \\
&=&\left( \begin{array}{ccc} Q_{11}(\tau,\mathbf{q}) & Q_{12}(\tau,\mathbf{q}) & Q^{\nu}_{13}(\tau,\mathbf{q}) \\ Q_{21}(\tau,\mathbf{q}) & Q_{22}(\tau,\mathbf{q}) & Q^{\nu}_{23}(\tau,\mathbf{q}) \\ Q^{\mu}_{31}(\tau,\mathbf{q}) & Q^{\mu}_{32}(\tau,\mathbf{q}) & Q^{\mu\nu}_{33}(\tau,\mathbf{q}) \end{array}\right)\left( \begin{array}{ccc} \Delta_{1\mathbf{q}} \\ \Delta_{2\mathbf{q}} \\A_{\nu\mathbf{q}} \end{array}\right).
\end{eqnarray}
The response functions $Q_{ij}$ are
\begin{eqnarray}
Q_{ij}(\tau-\tau',\mathbf{q})=-\sum_{\mathbf{p}\mathbf{p}'}\langle T_{\tau}[\bar{\Psi}_{\mathbf{p}}(\tau)\hat{\Sigma}_i\Psi_{\mathbf{p}+\mathbf{q}}(\tau)\bar{\Psi}_{\mathbf{p}'+\mathbf{q}}(\tau')\hat{\Sigma}_j\Psi_{\mathbf{p}'}(\tau')]\rangle.
\end{eqnarray}
Using a Fourier transform and making use of Wick's theorem, we obtain
\begin{eqnarray}\label{eqn:RF}
Q_{ij}(i\Omega_{l}, \mathbf{q})&=&\textrm{Tr}T\sum_{i\omega_n}\sum_{\mathbf{p}\mathbf{p}'} \hat{\Sigma}_i\hat{G}_{\mathbf{p}+\mathbf{q},\mathbf{p}'+\mathbf{q}}(i\omega_n+i\Omega_{l})\hat{\Sigma}_j\hat{G}_{\mathbf{p},\mathbf{p}'}(i\omega_n)\nonumber \\
&=&\textrm{Tr}T\sum_{i\omega_n}\sum_{\mathbf{p}}\hat{\Sigma}_i\hat{G}(P+Q,\mu)\hat{\Sigma}_j\hat{G}(P,\mu),
\end{eqnarray}
where $Q=(i\Omega_{l},\mathbf{q})$, $\Omega_{l}$ is the boson Matsubara frequency, $\hat{G}_{\mathbf{p},\mathbf{p}'}(i\omega_n)=\hat{G}_{\mathbf{p}}(i\omega_n)\delta_{\mathbf{p},\mathbf{p}'}$, and $\hat{G}_{\mathbf{p}}(i\omega_n)\equiv\hat{G}(P,\mu)=\frac{1}{i\omega_n-\hat{E}_{\mathbf{p}}}\gamma^0$. Inserting the above relations into Eq.~(\ref{eqn:RF}), the linear response matrix is given by
\begin{eqnarray}\label{eqn:RF1}
Q_{ij}(i\Omega_{l},\mathbf{q})=T\sum_{i\omega_n}\sum_{\mathbf{p}}\textrm{Tr}
\Big(\hat{\Sigma}_i\frac{1}{i\omega_n+i\Omega_{l}-\hat{E}_{\mathbf{p}+\mathbf{q}}}\gamma^0\hat{\Sigma}_j\frac{1}{i\omega_n-\hat{E}_{\mathbf{p}}}\gamma^0\Big).
\end{eqnarray}

Next we show that if the fluctuations of the order parameter are formulated as shown in Eq.~(\ref{eqn:H'}), our microscopic linear response theory is explicitly gauge invariant. Applying the condition $\eta_{1,2}=-\frac{2}{g}\Delta_{1,2}$ consistent with the gap equation to Eqs.~(\ref{eqn:Q}), we find
\begin{eqnarray}\label{D1D2}
& &\Delta_1=-\frac{Q^{\nu}_{13}\tilde{Q}_{22}-Q^{\nu}_{23}Q_{12}}{\tilde{Q}_{11}\tilde{Q}_{22}-Q_{12}Q_{21}}A_{\nu},\nonumber\\
& &\Delta_2=-\frac{Q^{\nu}_{23}\tilde{Q}_{11}-Q^{\nu}_{13}Q_{21}}{\tilde{Q}_{11}\tilde{Q}_{22}-Q_{12}Q_{21}}A_{\nu}.
\end{eqnarray}
where $\tilde{Q}_{11}\equiv \frac{2}{g}+Q_{11}$ and $\tilde{Q}_{22}\equiv \frac{2}{g}+Q_{22}$. After substituting the results into
\begin{eqnarray}
J^{\mu}\equiv \eta_3=Q^{\mu}_{31}\Delta_1+Q^{\mu}_{32}\Delta_2+Q^{\mu\nu}_{33}A_{\nu},
\end{eqnarray}
we get $J^{\mu}=K^{\mu\nu}A_{\nu}=(K^{\mu\nu}_0+\delta K^{\mu\nu})A_{\nu}$. Here $K^{\mu\nu}_0=Q^{\mu\nu}_{33}$ and
\begin{eqnarray}\label{dK}
\delta K^{\mu\nu}=-\frac{\tilde{Q}_{11}Q^{\mu}_{32}Q^{\nu}_{23}+\tilde{Q}_{22}Q^{\mu}_{31}Q^{\nu}_{13}-Q_{12}Q^{\mu}_{31}Q^{\nu}_{23}-Q_{21}Q^{\mu}_{32}Q^{\nu}_{13}}{\tilde{Q}_{11}\tilde{Q}_{22}-Q_{12}Q_{21}}.
\end{eqnarray}
The gauge invariance condition $q_{\mu}J^{\mu}=0$ leads to $q_{\mu}K^{\mu\nu}(Q)=0$, where $q_{\mu}\equiv Q=(i\Omega_{l},\mathbf{q})$ is the covariant form of the four-momentum. This condition is explicitly satisfied if the response functions satisfy the generalized Ward identities
\begin{eqnarray}\label{WI}
& &q_{\mu}Q^{\mu}_{31}=-2i\Delta Q_{21},\nonumber\\
& &q_{\mu}Q^{\mu}_{32}=-2i\Delta\tilde{Q}_{22},\nonumber\\
& &q_{\mu}Q^{\mu\nu}_{33}=-2i\Delta Q^{\nu}_{23}.
\end{eqnarray}
The derivations of these GWIs will be given in a moment. Firstly we show that the response functions indeed satisfy those GWIs. We observe that
\begin{eqnarray}
q_{\mu}K^{\mu\nu}&=&-2i\Delta Q^{\nu}_{23}+2i\Delta\frac{\tilde{Q}_{11}\tilde{Q}_{22}Q^{\nu}_{23}+\tilde{Q}_{22}Q_{21}Q^{\nu}_{13}-Q_{12}Q_{21}Q^{\nu}_{23}-Q_{21}\tilde{Q}_{22}Q^{\nu}_{13}}{\tilde{Q}_{11}\tilde{Q}_{22}-Q_{12}Q_{21}}\nonumber\\
&=&-2i\Delta Q^{\nu}_{23}+2i\Delta Q^{\nu}_{23}=0.
\end{eqnarray}

The proof of the GWIs (\ref{WI}) is sketched here. Our starting point is the expression (\ref{eqn:RF}). In what follows we will use the covariant form of the four-momentum $p_{\mu}\equiv P=(i \omega_n, \mathbf{p})$ and $P$ interchangeably. Moreover, $\sum_P\equiv T\sum_{i\omega_n}\sum_{\mathbf{p}}$. If we apply the analytical continuation $i\omega_n\rightarrow\omega+i\delta$, then $\sum_P=\int\frac{d^4P}{(2\pi)^4}$. By using the notation $\slashed{p}=\gamma^0i\omega_n-\vec{\gamma}\cdot\mathbf{p}$ we can express the bare and full propagators in Nambu space as
\begin{eqnarray}\label{PN1}
& &\hat{G}_0^{-1}(P,\mu)=\slashed{p}-m+\sigma_3\mu\gamma^0,\nonumber\\
& &\hat{G}^{-1}(P,\mu)=\hat{G}_0^{-1}(P,\mu)-\hat{\Sigma},
\end{eqnarray}
where the self-energy in Nambu space is $\hat{\Sigma}=-\Delta\sigma_1i\gamma_5$ and should not be confused with the effective vertex function $\hat{\mathbf{\Sigma}}$. Those expressions give
\begin{eqnarray}\label{eqn:G-G}
\sigma_3\hat{G}^{-1}(P+Q,\mu)-\hat{G}^{-1}(P,\mu)\sigma_3=\slashed{q}\sigma_3+2i\Delta\sigma_2i\gamma_5.
\end{eqnarray}
Eq.~(\ref{eqn:G-G}) will lead to the GWI (\ref{WI}). One can show that
\begin{eqnarray}\label{WI1}
& &q_{\mu}Q^{\mu}_{31}+2i\Delta Q_{21}\nonumber\\
&=&\textrm{Tr}\sum_P\big[\big(\sigma_3(\hat{G}^{-1}(P+Q,\mu)-\hat{G}^{-1}(P,\mu)\sigma_3\big)\hat{G}(P+Q,\mu)\sigma_1i\gamma_5\hat{G}(P,\mu)\big]\nonumber\\
&=&-2\textrm{Tr}\sum_P\big[\sigma_2\gamma_5\hat{G}(P,\mu)\big]\nonumber\\
&=&0,
\end{eqnarray}
where Eqs.~(\ref{app:DG}) has been used. Similarly, for the second GWI we can show that
\begin{eqnarray}
& &q_{\mu}Q^{\mu}_{32}+2i\Delta Q_{22}\nonumber\\
&=&\textrm{Tr}\sum_P\big[\sigma_3\sigma_2i\gamma_5\hat{G}(P,\mu)\big]-\textrm{Tr}\sum_P\big[\hat{G}(P+Q,\mu)i\gamma_5\sigma_2\sigma_3\big]\nonumber\\
&=&-2i\textrm{Tr}\sum_P\big[\sigma_1i\gamma_5\hat{G}(P,\mu)\big]\nonumber\\
&=&-\frac{4i}{g}\Delta,
\end{eqnarray}
where in the last line we have used Eq.~(\ref{eqn:n2}). Therefore we get the second GWI $q_{\mu}Q^{\mu}_{32}=-2i\Delta(Q_{22}+\frac{2}{g}\Delta)=-2i\Delta\tilde{Q}_{22}$. For the last GWI of Eq.~\eqref{WI}, we have
\begin{eqnarray}
& &q_{\mu}Q^{\mu\nu}_{33}+2i\Delta Q^{\nu}_{23}\nonumber\\
&=&\textrm{Tr}\sum_P\big[\gamma^{\nu}\hat{G}(P,\mu)\big]-\textrm{Tr}\sum_P\big[\hat{G}(P+Q,\mu)\gamma^{\nu}\big]\nonumber\\
&=&0.
\end{eqnarray}
This completes the proof.

\section{More about gauge invariance}\label{chap:5}
We have seen how the gauge invariance condition is satisfied by our response functions. Here we will clarify some subtleties from a generalized interaction picture. In our linear response theory, the Lagrangian density after the BCS approximation is given by
\begin{eqnarray}\label{LD}
\mathcal{L}_{BCS}&=&\mathcal{L}_{BCS0}+\mathcal{L}'\nonumber\\
&=&\bar{\Psi}\big(i\gamma^{\mu}\partial_{\mu}-m+\mu\gamma^{0}\sigma_3+\Delta i\gamma_5\sigma_1\big)\Psi-\bar{\Psi}\big(\Delta_1i\gamma_5\sigma_1+\Delta_2i\gamma_5\sigma_2+\slashed{A}\sigma_3\big)\Psi.
\end{eqnarray}
One can show that it is invariant under the generalized infinitesimal gauge transformation
\begin{eqnarray}\label{GT}
& &\Psi\rightarrow (1+i\sigma_3\chi)\Psi,\nonumber\\
& &\bar{\Psi}\rightarrow \bar{\Psi}(1-i\sigma_3\chi),\nonumber\\
& &\Delta\rightarrow\Delta,\nonumber\\
& &A_{\mu}\rightarrow A_{\mu}-\partial_{\mu}\chi,\nonumber\\
& &\Delta_1\rightarrow\Delta_1,\nonumber\\
& &\Delta_2\rightarrow\Delta_2+2\Delta\chi.
\end{eqnarray}
Under this transformation the two parts of the Lagrangian density transform as
\begin{eqnarray}
\mathcal{L}_{BCS0}&\rightarrow& \mathcal{L}_{BCS0}-\bar{\Psi}\sigma_3\slashed{\partial}\chi\Psi+i\chi\bar{\Psi}\Delta i\gamma_5[\sigma_1,\sigma_3]\Psi\nonumber\\
&=&\mathcal{L}_{BCS0}-\bar{\Psi}\sigma_3\slashed{\partial}\chi\Psi+2\chi\bar{\Psi}\Delta i\gamma_5\sigma_2\Psi, \nonumber\\
\mathcal{L}'&\rightarrow&\mathcal{L}'-2\chi\bar{\Psi}\Delta i\gamma_5\sigma_2\Psi+\bar{\Psi}\sigma_3\slashed{\partial}\chi\Psi.
\end{eqnarray}
Therefore $\mathcal{L}_{BCS}$ is invariant under the generalized infinitesimal gauge transformation (\ref{GT}). Here we emphasize that the transformation (\ref{GT}) only keeps terms linear in $\chi$ in the linear response theory.

The original mean-field Lagrangian density (\ref{LMF}) with a real $\Delta$ is not gauge invariant if the order parameter $\Delta$ is not perturbed by the gauge transformation. To circumvent this we assume that the effects are absorbed into the fluctuations of the order parameter $\Delta'$ while the equilibrium value of $\Delta$ is unchanged. From $\Delta(\mathbf{x})=g\langle\psi^{T}_{\uparrow}Ci\gamma_{5}\psi_{\downarrow}\rangle$ and $\Delta_2=\textrm{Im}\Delta'$ we see that only the imaginary part of the order parameter is perturbed by $\psi_{\uparrow\downarrow}\rightarrow(1+i\chi)\psi_{\uparrow\downarrow}$ while the (negative) real part is not, Therefore the perturbations from the gauge transformation are $\delta\Delta_1=0$ and $\delta\Delta_2=2i\chi\Delta$. It is the gauge transformation of $\Delta_2$ that cancels the term associated with $\Delta$ in the generalized gauge transformation and leads to the gauge invariance of the Lagrangian density.

As shown in Section.\ref{chap:3}, the perturbative Lagrangian density can be written as $\bar{\Psi}\hat{\mathbf{\Phi}}^T_{\mathbf{q}}\cdot\hat{\mathbf{\Sigma}}\Psi$, where $\hat{\mathbf{\Phi}}$ and $\hat{\mathbf{\Sigma}}$ are defined by Eq.~(\ref{GGP}). In fact $\hat{\mathbf{\Phi}}$ may be viewed as the generalized external gauge field and $\hat{\mathbf{\Sigma}}$ may be viewed as the generalized vertex function. In our theory, there are three different spaces: (i) the two-dimensional Nambu space where the Pauli matrices live, (ii) the four-dimensional representation space of the Clifford Algebra in which the $\gamma-$matrices live,  and (iii) a three-dimensional space which we will define as the generalized gauge space, where the generalized external potential $\hat{\mathbf{\Phi}}$ and generalized vertex function $\hat{\mathbf{\Sigma}}$ are defined.
Thus the transformation (\ref{GT}) is the corresponding generalized gauge transformation of the generalized external gauge field. Explicitly,
\begin{eqnarray} \label{GT2}
\hat{\mathbf{\Phi}}\rightarrow\hat{\mathbf{\Phi}}+\left(\begin{array}{c} 0 \\ 2\Delta\chi \\ -\partial_{\mu}\chi\end{array}\right).
\end{eqnarray}
In momentum space $-\partial_{\mu}\chi$ becomes $-iq_{\mu}\chi$. We define the generalized external momentum as $\hat{\mathbf{q}}\equiv(0,2i\Delta,q_{\mu})^T$ in the generalized gauge space. Then the generalized gauge transformation (\ref{GT2}) can be written as
\begin{eqnarray}
\hat{\mathbf{\Phi}}\rightarrow\hat{\mathbf{\Phi}}-i\hat{\mathbf{q}}\chi.
\end{eqnarray}
We saw that the Lagrangian density is invariant under the generalized gauge transformation (\ref{GT}) and now we want to find the corresponding GWI. By using the form of generalized external momentum, Eq.~(\ref{eqn:G-G}) becomes
\begin{eqnarray}\label{WI2}
\sigma_3\hat{G}^{-1}(P+Q,\mu)-\hat{G}^{-1}(P,\mu)\sigma_3=\slashed{q}\sigma_3+2i\Delta\sigma_2i\gamma_5=\hat{\mathbf{q}}^T\cdot{\hat{\mathbf{\Sigma}}}.
\end{eqnarray}
This is the GWI associated with the generalized gauge symmetry.

Next we adress what the conserved generalized current associated with this gauge transformation should be. Using the self-consistent condition $\eta_{1,2}=-\frac{2}{g}\Delta_{1,2}$, Eq.~(\ref{eqn:Q}) can be written as
\begin{eqnarray}\label{eqn:Q2}
\left( \begin{array}{c} 0 \\ 0 \\ J^{\mu} \end{array}\right)=\left( \begin{array}{ccc} \tilde{Q}_{11} & Q_{12} & Q^{\nu}_{13} \\ Q_{21} & \tilde{Q}_{22} & Q^{\nu}_{23} \\ Q^{\mu}_{31} & Q^{\mu}_{32} & Q^{\mu\nu}_{33} \end{array}\right)\left( \begin{array}{ccc} \Delta_{1} \\ \Delta_{2} \\A_{\nu} \end{array}\right).
\end{eqnarray}
We then define the generalized current $\hat{\mathbf{J}}\equiv(0,0,J^{\mu})^T$ and three generalized response-function vectors
\begin{eqnarray}\label{eqn:Q3}
\hat{\mathbf{Q}}_1=\left( \begin{array}{c} \tilde{Q}_{11} \\ Q_{21} \\ Q^{\mu}_{31} \end{array}\right),\qquad \hat{\mathbf{Q}}_2=\left( \begin{array}{c} Q_{12} \\ \tilde{Q}_{22} \\ Q^{\mu}_{32} \end{array}\right),\qquad \hat{\mathbf{Q}}^{\mu}_3=\left( \begin{array}{c} Q^{\mu}_{13} \\ Q^{\mu}_{23} \\ Q^{\mu\nu}_{33} \end{array}\right).
\end{eqnarray}
Then the generalized current (\ref{eqn:Q2}) becomes
\begin{eqnarray}
\hat{\mathbf{J}}=(\hat{\mathbf{Q}}_1,\hat{\mathbf{Q}}_2,\hat{\mathbf{Q}}^{\mu}_3)\cdot\hat{\mathbf{\Phi}},
\end{eqnarray}
The GWIs (\ref{WI}) for the response functions can also be written as
\begin{eqnarray}
\hat{\mathbf{q}}^T\cdot\hat{\mathbf{Q}}_i=0\quad\textrm{for $i=1,2,3$}.
\end{eqnarray}
Thus the GWIs directly lead to the conservation of the generalized current
\begin{eqnarray}
\hat{\mathbf{q}}^T\cdot\hat{\mathbf{J}}=(\hat{\mathbf{q}}^T\cdot\hat{\mathbf{Q}}_1,\hat{\mathbf{q}}^T\cdot\hat{\mathbf{Q}}_2,\hat{\mathbf{q}}^T\cdot\hat{\mathbf{Q}}^{\mu}_3)\cdot\hat{\mathbf{\Phi}}=0.
\end{eqnarray}
This gives
\begin{eqnarray}
q_{\mu}J^{\mu}=0.
\end{eqnarray}
Therefore $\hat{\mathbf{J}}$ is indeed the conserved current associated with the generalized gauge transformation. We see that the generalized gauge transformation leads to the usual $U(1)$ gauge invariance of our linear response theory. Importantly, the GWI (\ref{WI2}) for the generalized vertex function is exact since there are no high order corrections to the vertex in the linear response theory. Thus we have proved that our CFOP theory is indeed  gauge invariant.

\section{Expressions of the response functions}\label{chap:6}
It will greatly simplify our expressions of the response functions from the CFOP approach if we sort out the odevities of them first. Here we list the main  results and leave the details in Appdex.\ref{app:ob}. The odevity of the response functions about the four-momentum $Q=(i\Omega_{l},\mathbf{q})$ is
\begin{eqnarray}
Q_{ij}(i\Omega_{l},\mathbf{q})= (-1)^{(\delta^{2i}+\delta^{2j})}Q_{ij}(-i\Omega_{l},-\mathbf{q}).
\end{eqnarray}
The odevity about the gauge indices $i,j$ is given by
\begin{eqnarray}
Q_{ji}(i\Omega_{l},\mathbf{q})=(-1)^{(\delta^{2i}+\delta^{2j})}Q_{ij}(i\Omega_{l},\mathbf{q}).
\end{eqnarray}
The odevity about the boson Matsubara frequency $i\Omega_l$ is relatively complicated.
For $i=j=1,2$ we have
\begin{eqnarray}
Q_{ii}(i\Omega_{l},\mathbf{q})=Q_{ii}(-i\Omega_{l},\mathbf{q}).
\end{eqnarray}
For $i=j=3$  we have
\begin{eqnarray}
Q^{\mu\nu}_{33}(i\Omega_{l},\mathbf{q})=\left\{\begin{array}{cc} Q^{\mu\nu}_{33}(-i\Omega_{l},\mathbf{q}) & \textrm{if $\mu=\nu=0$ or $\mu=i$, $\nu=j$} \\ -Q^{\mu\nu}_{33}(-i\Omega_{l},\mathbf{q}) & \textrm{if $\mu=0$, $\nu=i$ or $\mu=i$, $\nu=0$}\end{array}\right.
\end{eqnarray}
For $i=1$, $j=2$ we have
\begin{eqnarray}
Q_{12}(i\Omega_{l},\mathbf{q})=-Q_{12}(-i\Omega_{l},\mathbf{q}).
\end{eqnarray}
For $i=1$, $j=3$ we have
\begin{eqnarray}
Q^{\mu}_{13}(i\Omega_{l},\mathbf{q})=\left\{\begin{array}{cc} Q^{\mu}_{13}(-i\Omega_{l},\mathbf{q}) & \textrm{if $\mu=0$}  \\ -Q^{\mu}_{13}(-i\Omega_{l},\mathbf{q}) & \textrm{if $\mu=i$ }\end{array}\right.
\end{eqnarray}
For $i=2$, $j=3$ we have
\begin{eqnarray}
Q^{\mu}_{23}(i\Omega_{l},\mathbf{q})=\left\{\begin{array}{cc} -Q^{\mu}_{23}(-i\Omega_{l},\mathbf{q}) & \textrm{if $\mu=0$}  \\ Q^{\mu}_{23}(-i\Omega_{l},\mathbf{q}) & \textrm{if $\mu=i$ }\end{array}\right.
\end{eqnarray}

After sorting out the odevities, the expressions of the response functions can be derived.
After summing the Matsubara frequencies, Eq.~(\ref{eqn:RF1}) becomes
\begin{eqnarray}\label{eqn:RF2}
Q_{ij}(i\Omega_{l},\mathbf{q})=\sum_{\mathbf{p}}\int d\epsilon_1\int d\epsilon_2\textrm{Tr}
\Big(\frac{f(\epsilon_1)-f(\epsilon_2)}{\epsilon_1-\epsilon_2-i\Omega_{l}}\hat{\Sigma}_i\delta(\epsilon_1-\hat{E}_{\mathbf{p}+\mathbf{q}})\gamma^0\hat{\Sigma}_j
\delta(\epsilon_2-\hat{E}_{\mathbf{p}})\gamma^0\Big).
\end{eqnarray}
The $\delta$-function operator can be decomposed as (see Appendix.\ref{app:C})
\begin{eqnarray}
\delta(\epsilon-\hat{E}_{\mathbf{p}})=\hat{u}^-_{\mathbf{p}}\delta(\epsilon-E^-_{\mathbf{p}})+\hat{v}^-_{\mathbf{p}}\delta(\epsilon+E^-_{\mathbf{p}})
+\hat{u}^+_{\mathbf{p}}\delta(\epsilon+E^+_{\mathbf{p}})+\hat{v}^+_{\mathbf{p}}\delta(\epsilon-E^+_{\mathbf{p}}).
\end{eqnarray}
We define the coherence coefficients as
\begin{eqnarray}\label{eqn:CEs}
& &(u^{\pm}u^{\pm})_{ij}=\textrm{Tr}[\hat{\Sigma}_i\hat{u}^{\pm}_{\mathbf{p}+\mathbf{q}}\gamma^0\hat{\Sigma}_j\hat{u}^{\pm}_{\mathbf{p}}\gamma^0],\mbox{ }(u^{\mp}u^{\pm})_{ij}=\textrm{Tr}[\hat{\Sigma}_i\hat{u}^{\mp}_{\mathbf{p}+\mathbf{q}}\gamma^0\hat{\Sigma}_j\hat{u}^{\pm}_{\mathbf{p}}\gamma^0],\nonumber\\
& &(u^{\pm}v^{\pm})_{ij}=\textrm{Tr}[\hat{\Sigma}_i\hat{u}^{\pm}_{\mathbf{p}+\mathbf{q}}\gamma^0\hat{\Sigma}_j\hat{v}^{\pm}_{\mathbf{p}}\gamma^0],\mbox{ }(u^{\mp}v^{\pm})_{ij}=\textrm{Tr}[\hat{\Sigma}_i\hat{u}^{\mp}_{\mathbf{p}+\mathbf{q}}\gamma^0\hat{\Sigma}_j\hat{v}^{\pm}_{\mathbf{p}}\gamma^0],\nonumber\\
& &(v^{\pm}u^{\pm})_{ij}=\textrm{Tr}[\hat{\Sigma}_i\hat{v}^{\pm}_{\mathbf{p}+\mathbf{q}}\gamma^0\hat{\Sigma}_j\hat{u}^{\pm}_{\mathbf{p}}\gamma^0],\mbox{ }(v^{\mp}u^{\pm})_{ij}=\textrm{Tr}[\hat{\Sigma}_i\hat{v}^{\mp}_{\mathbf{p}+\mathbf{q}}\gamma^0\hat{\Sigma}_j\hat{u}^{\pm}_{\mathbf{p}}\gamma^0],\nonumber\\
& &(v^{\pm}v^{\pm})_{ij}=\textrm{Tr}[\hat{\Sigma}_i\hat{v}^{\pm}_{\mathbf{p}+\mathbf{q}}\gamma^0\hat{\Sigma}_j\hat{v}^{\pm}_{\mathbf{p}}\gamma^0],\mbox{ }(v^{\mp}v^{\pm})_{ij}=\textrm{Tr}[\hat{\Sigma}_i\hat{v}^{\mp}_{\mathbf{p}+\mathbf{q}}\gamma^0\hat{\Sigma}_j\hat{v}^{\pm}_{\mathbf{p}}\gamma^0].
\end{eqnarray}
The response functions can be explicitly written down as
\begin{eqnarray}
& &Q_{ij}(i\Omega_{l},\mathbf{q})  \nonumber\\
=\sum_{\mathbf{p}}&\Big[& \frac{\big(f(E^-_{\mathbf{p}+\mathbf{q}})-f(E^-_{\mathbf{p}})\big)(u^-u^-)_{ij}}{E^-_{\mathbf{p}+\mathbf{q}}-E^-_{\mathbf{p}}-i\Omega_{l}}
-\frac{\big(1-f(E^-_{\mathbf{p}+\mathbf{q}})-f(E^-_{\mathbf{p}})\big)(u^-v^-)_{ij}}{E^-_{\mathbf{p}+\mathbf{q}}+E^-_{\mathbf{p}}-i\Omega_{l}}\nonumber\\
&-&\frac{\big(1-f(E^-_{\mathbf{p}+\mathbf{q}})-f(E^+_{\mathbf{p}})\big)(u^-u^+)_{ij}}{E^-_{\mathbf{p}+\mathbf{q}}+E^+_{\mathbf{p}}-i\Omega_{l}}
+\frac{\big(f(E^-_{\mathbf{p}+\mathbf{q}})-f(E^+_{\mathbf{p}})\big)(u^-v^+)_{ij}}{E^-_{\mathbf{p}+\mathbf{q}}-E^+_{\mathbf{p}}-i\Omega_{l}}\nonumber\\
&-&\frac{\big(1-f(E^-_{\mathbf{p}+\mathbf{q}})-f(E^-_{\mathbf{p}})\big)(v^-u^-)_{ij}}{E^-_{\mathbf{p}+\mathbf{q}}+E^-_{\mathbf{p}}+i\Omega_{l}}
+\frac{\big(f(E^-_{\mathbf{p}+\mathbf{q}})-f(E^-_{\mathbf{p}})\big)(v^-v^-)_{ij}}{E^-_{\mathbf{p}+\mathbf{q}}-E^-_{\mathbf{p}}+i\Omega_{l}}\nonumber\\
&+&\frac{\big(f(E^-_{\mathbf{p}+\mathbf{q}})-f(E^+_{\mathbf{p}})\big)(v^-u^+)_{ij}}{E^-_{\mathbf{p}+\mathbf{q}}-E^+_{\mathbf{p}}+i\Omega_{l}}
-\frac{\big(1-f(E^-_{\mathbf{p}+\mathbf{q}})-f(E^+_{\mathbf{p}})\big)(v^-v^+)_{ij}}{E^-_{\mathbf{p}+\mathbf{q}}+E^+_{\mathbf{p}}+i\Omega_{l}}\nonumber\\
&-&\frac{\big(1-f(E^+_{\mathbf{p}+\mathbf{q}})-f(E^-_{\mathbf{p}})\big)(u^+u^-)_{ij}}{E^+_{\mathbf{p}+\mathbf{q}}+E^-_{\mathbf{p}}+i\Omega_{l}}
+\frac{\big(f(E^+_{\mathbf{p}+\mathbf{q}})-f(E^-_{\mathbf{p}})\big)(u^+v^-)_{ij}}{E^+_{\mathbf{p}+\mathbf{q}}-E^-_{\mathbf{p}}+i\Omega_{l}}\nonumber\\
&+&\frac{\big(f(E^+_{\mathbf{p}+\mathbf{q}})-f(E^+_{\mathbf{p}})\big)(u^+u^+)_{ij}}{E^+_{\mathbf{p}+\mathbf{q}}-E^+_{\mathbf{p}}+i\Omega_{l}}
-\frac{\big(1-f(E^+_{\mathbf{p}+\mathbf{q}})-f(E^+_{\mathbf{p}})\big)(u^+v^+)_{ij}}{E^+_{\mathbf{p}+\mathbf{q}}+E^+_{\mathbf{p}}+i\Omega_{l}}\nonumber\\
&+&\frac{\big(f(E^+_{\mathbf{p}+\mathbf{q}})-f(E^-_{\mathbf{p}})\big)(v^+u^-)_{ij}}{E^+_{\mathbf{p}+\mathbf{q}}-E^-_{\mathbf{p}}-i\Omega_{l}}
-\frac{\big(1-f(E^+_{\mathbf{p}+\mathbf{q}})-f(E^-_{\mathbf{p}})\big)(v^+v^-)_{ij}}{E^+_{\mathbf{p}+\mathbf{q}}+E^-_{\mathbf{p}}-i\Omega_{l}}\nonumber\\
&-&\frac{\big(1-f(E^+_{\mathbf{p}+\mathbf{q}})-f(E^+_{\mathbf{p}})\big)(v^+u^+)_{ij}}{E^+_{\mathbf{p}+\mathbf{q}}+E^+_{\mathbf{p}}-i\Omega_{l}}
+\frac{\big(f(E^+_{\mathbf{p}+\mathbf{q}})-f(E^+_{\mathbf{p}})\big)(v^+v^+)_{ij}}{E^+_{\mathbf{p}+\mathbf{q}}-E^+_{\mathbf{p}}-i\Omega_{l}}\Big].
\end{eqnarray}
If $Q_{ij}$ is an even function of $i\Omega_{l}$, the expression reduces to
\begin{eqnarray}
& &Q_{ij}(i\Omega_{l},\mathbf{q})=\nonumber\\
\sum_{\mathbf{p}}&\Big[& \frac{\big(f(E^-_{\mathbf{p}+\mathbf{q}})-f(E^-_{\mathbf{p}})\big)\big(E^-_{\mathbf{p}+\mathbf{q}}-E^-_{\mathbf{p}}\big)
}{\big(E^-_{\mathbf{p}+\mathbf{q}}-E^-_{\mathbf{p}}\big)^2-(i\Omega_{l})^2}\big((u^-u^-)_{ij}+(v^-v^-)_{ij}\big)\nonumber\\
&-&\frac{\big(1-f(E^-_{\mathbf{p}+\mathbf{q}})-f(E^-_{\mathbf{p}})\big)\big(E^-_{\mathbf{p}+\mathbf{q}}+E^-_{\mathbf{p}}\big)}
{\big(E^-_{\mathbf{p}+\mathbf{q}}+E^-_{\mathbf{p}}\big)^2-(i\Omega_{l})^2}\big((u^-v^-)_{ij}+(v^-u^-)_{ij}\big)\nonumber\\
&-&\frac{\big(1-f(E^-_{\mathbf{p}+\mathbf{q}})-f(E^+_{\mathbf{p}})\big)\big(E^-_{\mathbf{p}+\mathbf{q}}+E^+_{\mathbf{p}}\big)}
{\big(E^-_{\mathbf{p}+\mathbf{q}}+E^+_{\mathbf{p}}\big)^2-(i\Omega_{l})^2}\big((u^-u^+)_{ij}+(v^-v^+)_{ij}\big)\nonumber\\
&+&\frac{\big(f(E^-_{\mathbf{p}+\mathbf{q}})-f(E^+_{\mathbf{p}})\big)\big(E^-_{\mathbf{p}+\mathbf{q}}-E^+_{\mathbf{p}}\big)
}{\big(E^-_{\mathbf{p}+\mathbf{q}}-E^+_{\mathbf{p}}\big)^2-(i\Omega_{l})^2}\big((u^-v^+)_{ij}+(v^-u^+)_{ij}\big)\nonumber\\
&+&\textrm{terms with super-indices $(+\leftrightarrow-)$}\Big].
\end{eqnarray}
If $Q_{ij}$ is an odd function of $i\Omega_{l}$, the expression reduces to
\begin{eqnarray}
& &Q_{ij}(i\Omega_{l},\mathbf{q})=\nonumber\\
i\Omega_{l}\sum_{\mathbf{p}}&\Big[& \frac{f(E^-_{\mathbf{p}+\mathbf{q}})-f(E^-_{\mathbf{p}})
}{\big(E^-_{\mathbf{p}+\mathbf{q}}-E^-_{\mathbf{p}}\big)^2-(i\Omega_{l})^2}\big((u^-u^-)_{ij}-(v^-v^-)_{ij}\big)\nonumber\\
&-&\frac{1-f(E^-_{\mathbf{p}+\mathbf{q}})-f(E^-_{\mathbf{p}})}
{\big(E^-_{\mathbf{p}+\mathbf{q}}+E^-_{\mathbf{p}}\big)^2-(i\Omega_{l})^2}\big((u^-v^-)_{ij}-(v^-u^-)_{ij}\big)\nonumber\\
&-&\frac{1-f(E^-_{\mathbf{p}+\mathbf{q}})-f(E^+_{\mathbf{p}})}
{\big(E^-_{\mathbf{p}+\mathbf{q}}+E^+_{\mathbf{p}}\big)^2-(i\Omega_{l})^2}\big((u^-u^+)_{ij}-(v^-v^+)_{ij}\big)\nonumber\\
&+&\frac{f(E^-_{\mathbf{p}+\mathbf{q}})-f(E^+_{\mathbf{p}})
}{\big(E^-_{\mathbf{p}+\mathbf{q}}-E^+_{\mathbf{p}}\big)^2-(i\Omega_{l})^2}\big((u^-v^+)_{ij}-(v^-u^+)_{ij}\big)\nonumber\\
&-&\textrm{terms with super-indices $(+\leftrightarrow-)$}\Big].
\end{eqnarray}
The coherence coefficients are listed in Appendix.\ref{app:CE}.

\section{Collective mode contribution to response functions}\label{chap:coll}
In Section \ref{chap:3} one saw that the conservation of the induced current $q_{\mu}J^{\mu}=0$ requires $q_{\mu}K'^{\mu\nu}(Q)A'_{\nu}=0$, where  $A'_{\mu}$ and $K'^{\mu\nu}(Q)$ are an external field and its corresponding response kernel. In general, if one applies a gauge transformation $A_{\mu}=A'_{\mu}-iq_{\mu}\chi$, the response kernel transforms as $K'^{\mu\nu}(Q)\rightarrow K^{\mu\nu}(Q)$ and the gauge invariant condition requires $q_{\mu}K^{\mu\nu}(Q)A_{\nu}=0$. From the gauge invariance condition $q_{\mu}K'^{\mu\nu}_0(Q)A'_{\nu}=0$, one obtains $i\chi=-\frac{q_{\mu}K'^{\mu\nu}A_{\nu}}{q_{\alpha}K'^{\alpha\beta}q_{\beta}}$. Using the gauge invariance condition again one obtains
\begin{eqnarray}
K^{\mu\nu}=K'^{\mu\nu}-\frac{K'^{\mu\lambda}q_{\lambda}q_{\delta}K'^{\delta\nu}}{q_{\alpha}K'^{\alpha\beta}q_{\beta}}.
\end{eqnarray}

This is the general expression of how the response kernel transforms under gauge transformations. Clearly, the zeros of $q_{\mu}K'^{\mu\nu}q_{\nu}=0$, if exist, indicate the presence of collective excitations. In the following we will show that the massless Nambu-Goldstone boson indeed contributes. We simplify the expression of $K^{\mu\nu}$ by the response functions we obtained so far. Following the discussions in Ref.~\cite{KosztinPRB00}, we rewrite $K^{\mu\nu}$ in a more compact form
\begin{eqnarray}
K^{\mu\nu}=K'^{\mu\nu}_0+\delta K'^{\mu\nu}
\end{eqnarray}
with
\begin{eqnarray}
K'^{\mu\nu}_0=Q^{\mu\nu}_{33}-\frac{Q^{\mu}_{31}Q^{\nu}_{13}}{\tilde{Q}_{11}},\quad \delta K'^{\mu\nu}=-\frac{Q'^{\mu}_{32}Q'^{\nu}_{23}}{\tilde{Q}'_{22}},
\end{eqnarray}
where
\begin{eqnarray}
Q'^{\mu}_{32}=Q^{\mu}_{32}-\frac{Q_{12}}{\tilde{Q}_{11}}Q^{\mu}_{31},\quad \tilde{Q}'_{22}=\tilde{Q}_{22}-\frac{Q_{12}Q_{21}}{\tilde{Q}_{11}}.
\end{eqnarray}
By using the GWI (\ref{WI}) and the symmetry (\ref{D11}), one can show that
\begin{eqnarray}
\delta K'^{\mu\nu}=-\frac{K'^{\mu\lambda}_0q_{\lambda}q_{\delta}K'^{\delta\nu}_0}{q_{\alpha}K'^{\alpha\beta}_0q_{\beta}}.
\end{eqnarray}
This implies that $K^{\mu\nu}$ can be thought of as a functional of $K^{\mu\nu}_0$ and the denominator of the second term is expressed as $q_{\mu}K'^{\mu\nu}_0q_{\nu}=4\Delta^2\tilde{Q}'_{22}$. This expression is useful in the analysis of the dispersion of collective modes because an expansion around a small four-momentum around the pole of $K^{\mu\nu}$ gives the dispersion of the Goldstone boson. We suggest that it is the contribution of the Goldstone boson that leads the kernel to respect the gauge invariance condition. After making the analytical continuation $i\Omega_{l}\rightarrow\Omega+i\delta$, the dispersion of the collective mode at $T=0$ is evaluated wtih $\Omega\rightarrow 0$ and $\mathbf{q}\rightarrow\mathbf{0}$. Note that $\mathbf{q}\rightarrow\mathbf{0}$ does not imply that the system is in the nonrelativistic limit. One necessary condition for that limit is $k_F\ll m$.

\subsection{Nonrelativistic Limit}
In this limit, all contributions from the negative energy states vanish since $\Lambda_-(\mathbf{p})\simeq 0$.
Therefore, we have
\begin{eqnarray}\label{tmp3}
& &\tilde{Q}_{22}(\omega,\mathbf{q})=\sum_{\mathbf{p}}\frac{\omega^2-(\xi^-_{\mathbf{p}+\mathbf{q}}-\xi^-_{\mathbf{p}})^2}{E^-_{\mathbf{p}+\mathbf{q}}E^-_{\mathbf{p}}}\frac{E^-_{\mathbf{p}+\mathbf{q}}+E^-_{\mathbf{p}}}{\omega^2-(E^-_{\mathbf{p}+\mathbf{q}}+E^-_{\mathbf{p}})^2}\frac{B(\mathbf{p},\mathbf{q})}{2\epsilon_{\mathbf{p}+\mathbf{q}}\epsilon_{\mathbf{p}}},
\end{eqnarray}
where $B(\mathbf{p},\mathbf{q})=\epsilon_{\mathbf{p}+\mathbf{q}}\epsilon_{\mathbf{p}}+\epsilon^2_{\mathbf{p}}+\mathbf{p}\cdot\mathbf{q}$ is defined in Appendix.\ref{app:CE} and we have omitted the term proportional to $A(\mathbf{p},\mathbf{q})$ since $\lim_{\mathbf{q}\rightarrow{0}}A(\mathbf{p},\mathbf{q})=0$. Due to the particle-hole symmetry, $Q_{12}$ vanishes identically. Therefore $\tilde{Q}'_{22}=\tilde{Q}_{22}$ so we only need to find a solution to $\tilde{Q}_{22}=0$. As $\mathbf{q}\rightarrow\mathbf{0}$, $\xi^-_{\mathbf{p}}\simeq \frac{\mathbf{p}^2}{2m}-(\mu-m)=\frac{\mathbf{p}^2}{2m}-\mu^-$, where $\mu^-$ plays the role of the conventional nonrelativistic chemical potential.

At zero temperature, we keep the lowest order terms of $\omega$ and $\mathbf{q}$ of $\tilde{Q}_{22}$ and it becomes
\begin{eqnarray}
& &\tilde{Q}_{22}(\omega,\mathbf{q})\simeq-\frac{N(0)}{2}\int^{+\infty}_{-\infty}d\xi^-_{\mathbf{p}}\int^1_{-1}d\textrm{cos}\theta\frac{\omega^2-\frac{q^2p^2\textrm{cos}^2\theta}{m^2}}{E^{-2}_{\mathbf{p}}}\frac{1}{2E^{-}_{\mathbf{p}}}\nonumber \\
&=&-\frac{N(0)}{2}\int^{+\infty}_{-\infty}d\xi^-_{\mathbf{p}}\Big(\frac{\omega^2}{E^{-3}_{\mathbf{p}}}-\frac{1}{3}\frac{q^2p^2}{m^2}\frac{1}{E^{-3}_{\mathbf{p}}}\Big)\nonumber \\
&=&-\frac{N(0)}{\Delta^2}\big(\omega^2-\frac{2}{3}\frac{q^2\mu^-}{m}\big),
\end{eqnarray}
where we have used $p^2=2m(\xi^-_{\mathbf{p}}+\mu^-)$ and $N(0)$ is the density of states near the Fermi surface. Note that $\mu^-\simeq\epsilon_F=\frac{k^2_F}{2m}$ in BCS theory so we found
\begin{eqnarray}  \label{eqn:Q22}
& &\tilde{Q}_{22}(\omega,\mathbf{q})\simeq-\frac{N(0)}{\Delta^2}\big(\omega^2-\frac{1}{3}\frac{q^2k^2_F}{m^2}\big)=-\frac{N(0)}{\Delta^2}\big(\omega^2-c^2_sq^2\big),
\end{eqnarray}
where $c_s=\frac{1}{\sqrt{3}}\frac{k_F}{m}=\frac{1}{\sqrt{3}}v_F$ is the sound speed of a BCS superfluid. Thus $\tilde{Q}_{22}(0,\mathbf{0})=0$ indeed and the expansion shows the dispersion $\omega=c_sq$ of the massless Goldstone boson.

\subsection{Relativistic Limit}
Next we consider an ultra-relativistic BCS superfluid which is characterized by $k_F\gg m$ and $\Delta\ll\mu\simeq\epsilon_F=\sqrt{k^2_F+m^2}\simeq k_F$. Again the anti-fermion contribution can be safely ignored and due to the particle-hole symmetry we have $Q_{12}=0$. Note that $\xi^-_{\mathbf{p}+\mathbf{q}}\simeq\xi^-_{\mathbf{p}}+\mathbf{q}\cdot\nabla\xi^-_{\mathbf{p}}=\xi^-_{\mathbf{p}}+\frac{\mathbf{q}\cdot\mathbf{p}}{\epsilon_{\mathbf{p}}}$. Therefore we have
\begin{eqnarray}
& &\tilde{Q}_{22}(\omega,\mathbf{q})\simeq-\frac{1}{4\pi^2}\int^{+\infty}_{0}dpp^2\int^1_{-1}d\textrm{cos}\theta\frac{\omega^2-\frac{q^2p^2\textrm{cos}^2\theta}{\epsilon_{\mathbf{p}}^2}}{E^{-2}_{\mathbf{p}}}\frac{1}{2E^{-}_{\mathbf{p}}}\nonumber \\
&=&-\frac{1}{4\pi^2}\int^{+\infty}_{0}dpp^2\Big(\frac{\omega^2}{E^{-3}_{\mathbf{p}}}-\frac{1}{3}\frac{q^2p^2}{\epsilon_{\mathbf{p}}^2}\frac{1}{E^{-3}_{\mathbf{p}}}\Big)\nonumber \\
&=&-\frac{1}{4\pi^2}\int^{+\infty}_{0}dpp^2\Big(\frac{1}{E^{-3}_{\mathbf{p}}}(\omega^2-\frac{q^2}{3})+\frac{1}{3}\frac{q^2m^2}{\epsilon_{\mathbf{p}}^2}\frac{1}{E^{-3}_{\mathbf{p}}}\Big),
\end{eqnarray}
where $p^2=\epsilon^2_{\mathbf{p}}-m^2$ has been used. The second term in the big bracket is at most of the leading order of $\frac{m^2}{\mu^3}$. Since $m\ll k_F\simeq\mu$, it can be ignored. Therefore $\tilde{Q}_{22}(0,\mathbf{0})=0$ and the expansion of $\tilde{Q}_{22}(\omega,\mathbf{q})$ leads to $\omega=\frac{1}{\sqrt{3}}q$ so the contribution from the massless Goldstone boson is clearly demonstrated. The dispersion also implies $c_s=\frac{1}{\sqrt{3}}$, which is a well-known result for ultra-relativistic BCS theory.

\section{Compressibility from response function}\label{chap:dndmu}
The isothermal compressibility is given by $\kappa\equiv n^{-2}(\partial n/\partial\mu)$. Here the density susceptibility can be inferred from the response functions by \cite{Kubo_book}
\begin{equation}\label{eq:dndmudef}
\frac{\partial n}{\partial \mu}=-K^{00}(\omega=0,\mathbf{q}\rightarrow\mathbf{0}).
\end{equation}
At $T=0$, the number equation (\ref{ENUM}) and the gap equation (\ref{EGAP}) are
\begin{eqnarray}
& &n=2\sum_{\mathbf{p}}\big(\frac{\xi^+_{\mathbf{p}}}{E^+_{\mathbf{p}}}-\frac{\xi^-_{\mathbf{p}}}{E^-_{\mathbf{p}}}\big),\\
& &\frac{1}{g}=\sum_{\mathbf{p}}\big(\frac{1}{E^-_{\mathbf{p}}}+\frac{1}{E^+_{\mathbf{p}}}\big).
\end{eqnarray}
The density susceptibility can be obtained from these two equations. We treat $\Delta$ as a function $\Delta(\mu)$ of $\mu$. Differentiating the gap equation with respect to $\mu$ one obtains
\begin{eqnarray}
\frac{\partial\Delta}{\partial\mu}=\frac{\sum_{\mathbf{p}}\big(\frac{\xi^-_{\mathbf{p}}}{E^{-3}_{\mathbf{p}}}-
\frac{\xi^+_{\mathbf{p}}}{E^{+3}_{\mathbf{p}}}\big)}{\Delta\sum_{\mathbf{p}}\big(\frac{1}{E^{-3}_{\mathbf{p}}}+
\frac{1}{E^{+3}_{\mathbf{p}}}\big)}.
\end{eqnarray}
Differentiating the number equation with respect to $\mu$ and using the above result, one gets
\begin{eqnarray}\label{DSus}
\frac{\partial n}{\partial\mu}
=2\Delta^2\sum_{\mathbf{p}}\Big(\frac{1}{E^{-3}_{\mathbf{p}}}+\frac{1}{E^{+3}_{\mathbf{p}}}\Big)+\frac{2\Big[\sum_{\mathbf{p}}\big(\frac{\xi^-_{\mathbf{p}}}{E^{-3}_{\mathbf{p}}}-
\frac{\xi^+_{\mathbf{p}}}{E^{+3}_{\mathbf{p}}}\big)\Big]^2}{\sum_{\mathbf{p}}\big(\frac{1}{E^{-3}_{\mathbf{p}}}+
\frac{1}{E^{+3}_{\mathbf{p}}}\big)}.
\end{eqnarray}

Next we check if $K^{00}(0,\mathbf{q}\rightarrow\mathbf{0})$ can give the same density susceptibility. When $\omega=0$, $Q^0_{23}=Q^0_{32}=Q_{12}=Q_{21}=0$. Therefore from Eq.(\ref{dK}) we have
\begin{eqnarray}
K^{00}(0,\mathbf{q})=Q^{00}_{33}(0,\mathbf{q})-\frac{Q^0_{13}(0,\mathbf{q})Q^0_{31}(0,\mathbf{q})}{\tilde{Q}_{11}(0,\mathbf{q})}.
\end{eqnarray}
In the limit $\mathbf{q}\rightarrow\mathbf{0}$, one gets $B(\mathbf{p},\mathbf{0})=2$ and $A(\mathbf{p},\mathbf{0})=0$. Then at $T=0$ one has
\begin{eqnarray}
Q^{00}_{33}(0,\mathbf{q}\rightarrow\mathbf{0})&=&-2\Delta^2\sum_{\mathbf{p}}\Big(\frac{1}{E^{-3}_{\mathbf{p}}}+\frac{1}{E^{+3}_{\mathbf{p}}}\Big),\nonumber\\
Q^{0}_{13}(0,\mathbf{q}\rightarrow\mathbf{0})&=&Q^{0}_{31}(0,\mathbf{q}\rightarrow\mathbf{0})=-2\Delta\sum_{\mathbf{p}}\Big(\frac{\xi^-_{\mathbf{p}}}{E^{-3}_{\mathbf{p}}}-
\frac{\xi^+_{\mathbf{p}}}{E^{+3}_{\mathbf{p}}}\Big),\nonumber\\
\tilde{Q}_{11}(0,\mathbf{q}\rightarrow\mathbf{0})&=&\frac{2}{g}+Q_{11}(0,\mathbf{q}\rightarrow\mathbf{0})\nonumber\\
&=&2\sum_{\mathbf{p}}\Big(\frac{1}{E^-_{\mathbf{p}}}+\frac{1}{E^+_{\mathbf{p}}}\Big)
-2\sum_{\mathbf{p}}\Big(\frac{\xi^{-2}_{\mathbf{p}}}{E^{-3}_{\mathbf{p}}}+\frac{\xi^{+3}_{\mathbf{p}}}{E^{+3}_{\mathbf{p}}}\Big)\nonumber\\
&=&2\Delta^2\sum_{\mathbf{p}}\Big(\frac{1}{E^{-3}_{\mathbf{p}}}+\frac{1}{E^{+3}_{\mathbf{p}}}\Big).
\end{eqnarray}
After comparing this with Eq.~(\ref{DSus}), one finds that Eq.~\eqref{eq:dndmudef} is indeed satisfied. This consistency implies that the approximation of the fermion propagator is compatible with that of the response functions. Thus we emphasize that the collective-mode contribution is important in maintaining the integrity of the CFOP formalism.

\section{Meissner effect and superfluid density}\label{chap:Meissner}
The Meissner effect can be demonstrated by examining the behavior of the response kernel $\tensor{K}^{ij}(0,\mathbf{q})$ as $\mathbf{q}\rightarrow\mathbf{0}$. In the previous section, we have learned that $\tensor{K}^{ij}=\tensor{K}_0^{ij}+\delta \tensor{K}^{ij}$, where
$\tensor{K}_0^{ij}=\tensor{Q}^{ij}_{33}$ and
\begin{eqnarray}
\delta \tensor{K}^{ij}=-\frac{\tilde{Q}_{11}\mathbf{Q}^i_{32}\mathbf{Q}^j_{23}+\tilde{Q}_{22}\mathbf{Q}^i_{31}\mathbf{Q}^j_{13}-2Q_{12}\mathbf{Q}^i_{31}\mathbf{Q}^j_{23}}{\tilde{Q}_{11}\tilde{Q}_{22}-Q_{12}Q_{21}}
\end{eqnarray}
denoting the contribution from collective modes. However, in our model we found that collective mode effects do not contribute to the transverse components of the response functions in this limit and should not affect the Meissner effect. This is verified as the following. A tensor $\tensor{P}^{ij}$ can always be decomposed into the longitudinal and the transverse parts $P_{L}$ and $P_T$, where $P_{L}=\hat{\mathbf{q}}\cdot\tensor{P}\cdot\hat{\mathbf{q}}$ and $P_T=(\sum_i\tensor{P}^{ii}-P_L)/2$. Assuming that $\mathbf{q}$ is parallel to the $z$-axis, in the limit $\mathbf{q}\rightarrow\mathbf{0}$ among all components of the response functions only $\mathbf{Q}^z_{31}$ and $\mathbf{Q}^z_{32}$ do not vanish to the first order of $q$. From this we conclude that $\lim_{\mathbf{q}\rightarrow\mathbf{0}}\mathbf{Q}_{3i}\cdot\mathbf{Q}_{3j}=\lim_{\mathbf{q}\rightarrow\mathbf{0}}\hat{\mathbf{q}}\cdot\mathbf{Q}_{3i}\mathbf{Q}_{3j}\cdot\hat{\mathbf{q}}$. This means that the transverse component of the tensor $\mathbf{Q}_{3i}\mathbf{Q}_{3j}$ vanishes in the limit $\mathbf{q}\rightarrow\mathbf{0}$. Therefore the transverse part of $\tensor{K}$ receives no contribution from the collective modes so we only need to consider $\tensor{K}_0^{ij}=\tensor{Q}^{ij}_{33}$ in the study of the Meissner effect. Defining $\hat{\mathbf{p}}\equiv\mathbf{p}/\epsilon_{\mathbf{p}}$, the longitudinal and transverse parts of the response functions are given by
\begin{eqnarray}
& &\lim_{\mathbf{q}\rightarrow\mathbf{0}}\tensor{K}_{0L}^{ij}(0,\mathbf{q})=4\int\frac{d^{3}\mathbf{p}}{(2\pi)^{3}}\hat{p}^{i}\hat{p}^{j}\Big(\frac{\partial f(E^{-}_{\mathbf{p}})}{\partial E^{-}_{\mathbf{p}}}+\frac{\partial f(E^{+}_{\mathbf{p}})}{\partial E^{+}_{\mathbf{p}}}\Big),\\
& &\lim_{\mathbf{q}\rightarrow\mathbf{0}}\tensor{K}_{0T}^{ij}(0,\mathbf{q})=4\int\frac{d^{3}\mathbf{p}}{(2\pi)^{3}}(\delta^{ij}-\hat{p}^{i}\hat{p}^{j})\Big[
\big(1+\frac{\xi^{+}_{\mathbf{p}}\xi^{-}_{\mathbf{p}}-\Delta^{2}}{E^{+}_{\mathbf{p}}E^{-}_{\mathbf{p}}}\big)\frac{f(E^{-}_{\mathbf{p}})+f(E^{+}_{\mathbf{p}})-1}{E^{-}_{\mathbf{p}}+E^{+}_{\mathbf{p}}}\nonumber\\
& &\qquad\qquad\qquad\qquad\qquad\qquad\qquad\qquad+\big(1-\frac{\xi^{+}_{\mathbf{p}}\xi^{-}_{\mathbf{p}}-\Delta^{2}}{E^{+}_{\mathbf{p}}E^{-}_{\mathbf{p}}}\big)\frac{f(E^{-}_{\mathbf{p}})-f(E^{+}_{\mathbf{p}})}{E^{-}_{\mathbf{p}}-E^{+}_{\mathbf{p}}}\Big]
\end{eqnarray}
Now we focus on the nonrelativistic BCS limit where $\xi^+_{\mathbf{p}}\simeq E^+_{\mathbf{p}}\simeq 2m$ (since $\mu\simeq m$), $\epsilon_{\mathbf{p}}\simeq m$ and $\mu^-,\Delta\ll m$. Here a suitable regularization is needed to give physical results. This is done by subtracting the vacuum contribution from $K^{ij}$, i.e., $K^{ij}(Q)\rightarrow K^{ij}(Q)-K^{ij}(Q)|_{T=\Delta=0,\mu=m}$. Finally we have
\begin{eqnarray}
\lim_{\mathbf{q}\rightarrow\mathbf{0}}\tensor{K}_{0L}^{ij}(0,\mathbf{q})&=&4\int\frac{d^{3}\mathbf{p}}{(2\pi)^{3}}\frac{\mathbf{p}^i\mathbf{p}^j}{m^2}\frac{\partial f(E^{-}_{\mathbf{p}})}{\partial E^{-}_{\mathbf{p}}}+O(\frac{1}{m^3}),\\
\lim_{\mathbf{q}\rightarrow\mathbf{0}}\tensor{K}_{0T}^{ij}(0,\mathbf{q})&=&4\frac{\delta^{ij}}{m}\int\frac{d^{3}\mathbf{p}}{(2\pi)^{3}}\Big[u^{-2}_{\mathbf{p}}f(E^-_{\mathbf{p}})+v^{-2}_{\mathbf{p}}f(-E^-_{\mathbf{p}})\Big]+O(\frac{1}{m^3})\nonumber\\
&=&2\frac{\delta^{ij}}{m}n^{\textrm{NR}}+O(\frac{1}{m^3}),
\end{eqnarray}
where $n^{\textrm{NR}}$ is the fermion number in the nonrelativistic limit. Therefore, from $J^{i}=K^{ij}A_j$ we get the well-known London equation
\begin{eqnarray}
\mathbf{J}(\mathbf{q})&=&-\frac{2}{m}\Big(\mathbf{A}(\mathbf{q})n^{\textrm{NR}}+\frac{2}{m}\int\frac{d^{3}\mathbf{p}}{(2\pi)^{3}}\mathbf{p}\big(\mathbf{p}\cdot\mathbf{A}(\mathbf{q})\big)\frac{\partial f(E^{-}_{\mathbf{p}})}{\partial E^{-}_{\mathbf{p}}}\Big)\nonumber\\
&=&-\frac{2}{m}\mathbf{A}(\mathbf{q})n^{\textrm{NR}}\Big(1+\frac{1}{3\pi^2mn^{\textrm{NR}}}\int_0^{+\infty}dpp^4\frac{\partial f(E^{-}_{\mathbf{p}})}{\partial E^{-}_{\mathbf{p}}}\Big)\nonumber\\
&=&-\frac{2}{m}\mathbf{A}(\mathbf{q})n_s,
\end{eqnarray}
where
\begin{eqnarray}
n_s=n^{\textrm{NR}}-\frac{1}{3\pi^2mn^{\textrm{NR}}}\int_0^{+\infty}dpp^4\big(-\frac{\partial f(E^{-}_{\mathbf{p}})}{\partial E^{-}_{\mathbf{p}}}\big)
\end{eqnarray}
is the nonrelativistic superfluid density. Again the extra factor $2$ in the London equation comes from the fact that we introduce the pseudo-spin $\sigma=\uparrow$, $\downarrow$ so there are two times more degrees of freedom.

\section{Nambu's integral equation for relativistic BCS superfluids}\label{chap:Nambu}
Before closing our discussions on the CFOP theory of relativistic BCS superfluids, we present a generalization of Nambu's integral-equation approach. For nonrelativistic BCS superfluids, the spontaneously broken $U(1)$ symmetry can be restored in the linear response theory by Nambu's approach \cite{Nambu60}. In conventional BCS theory, the self-energy of the fermion is approximated by an integral equation which consists of a ladder approximation for the electron-phonon interaction. Nambu proposed that the EM vertex in the linear response theory should be corrected in the same way as the self energy. Hence the EM vertex function also follows the integral equation
\begin{eqnarray}
\hat{\Gamma}^{\textrm{NR}}_{\mu}(P+Q,P)&=&\hat{\gamma}^{\textrm{NR}}_{\mu}(P+Q,P)- \nonumber \\
& &g^{\textrm{NR}}\sum_K\sigma_3\hat{G}^{\textrm{NR}}(K+Q)\hat{\Gamma}^{\textrm{NR}}_{\mu}(K+Q,K)\hat{G}^{\textrm{NR}}(K)\sigma_3,
\end{eqnarray}
where the superscript ``NR'' denotes the corresponding nonrelativistic quantities. Explicitly, the solution to this equation should be an EM vertex that respects the GWIs. For the relativistic BCS model, following similar arguments we will derive the corresponding integral equation for the EM vertex.

In Section {\ref{chap:3}}, we found that the bare polarization function $K^{\mu\nu}_0=Q^{\mu\nu}_{33}$ does not satisfy the gauge-invariance condition $q_{\mu}K^{\mu\nu}_0=0$. This is because the collective modes which correspond to the fluctuations of the order parameter also contribute to response functions. The EM vertex without collective-mode effects in Nambu space is given by
\begin{eqnarray}
\hat{\gamma}^{\mu}=\sigma_3\gamma^{\mu}=\left(\begin{array}{cc} \gamma^{\mu} & 0 \\ 0 & -\gamma^{\mu}\end{array}\right).
\end{eqnarray}
Then the bare polarization function (see Eq.~(\ref{eqn:RF1})) can be written as
\begin{eqnarray}\label{K0}
K^{\mu\nu}_0(Q)=\textrm{Tr}\sum_P\big(\hat{\gamma}^{\mu}\hat{G}(P+Q,\mu)\hat{\gamma}^{\nu}\hat{G}(P,\mu)\big).
\end{eqnarray}
The violation of the conservation law can be traced back to the use of the \textit{full} fermion propagator and the \textit{bare} EM vertex simultaneously \cite{Schrieffer_book,ReddyPRC04} so the approximations of the fermion propagator and EM vertex are not treated on equal footing. In quantum electrodynamics, the gauge invariance, or equivalently the Ward identity, can be maintained order by order. However, the generalized Ward identity is not respected even at the tree level for both relativistic and nonrelativistic BCS models if the approximations for the vertex and the self energy are different.

For the bare EM vertex $\hat{\gamma}^{\mu}$ in the relativistic BCS theory, one has
\begin{eqnarray}\label{FWI0}
q_{\mu}\hat{\gamma}^{\mu}(Q)=\slashed{q}\sigma_3=\sigma_3\hat{G}_0^{-1}(P+Q,\mu)-\hat{G}_0^{-1}(P,\mu)\sigma_3.
\end{eqnarray}
Note that the bare propagator appears in the identity. Thus a gauge invariant EM vertex must satisfy
\begin{eqnarray}\label{FWI1}
q_{\mu}\hat{\Gamma}^{\mu}(Q)=\sigma_3\hat{G}^{-1}(P+Q,\mu)-\hat{G}^{-1}(P,\mu)\sigma_3.
\end{eqnarray}
If we define the correction of the EM vertex as $\hat{\Gamma}^{\mu}(Q)-\hat{\gamma}^{\mu}(Q)=\delta\hat{\Gamma}^{\mu}(Q)$, then from the two equations above and Eqs.~(\ref{PN1}), one can get the GWI which is associated with the self energy:
\begin{eqnarray}\label{FWI2}
q_{\mu}\delta\hat{\Gamma}^{\mu}(Q)=\hat{\Sigma}\sigma_3-\sigma_3\hat{\Sigma}=2i\Delta\sigma_2i\gamma_5.
\end{eqnarray}
One possible method to respect the GWI (\ref{FWI1}) or (\ref{FWI2}), as pointed out by Nambu \cite{Nambu60}, is to treat the full EM vertex $\hat{\Gamma}^{\mu}$ in the same way as how the self energy is approximated. That is, the full EM vertex of relativistic BCS superfluids should satisfy the integral equation
\begin{eqnarray}\label{FD1}
\hat{\Gamma}^{\mu}(Q)&=&\hat{\gamma}^{\mu}(Q)-2g\sum_P\sigma_3\Big[\hat{G}(P,\mu)\hat{\Gamma}^{\mu}(Q)\hat{G}(P+Q,\mu)+\hat{G}(P,-\mu)\hat{\Gamma}^{\mu}(Q)\hat{G}(P+Q,-\mu)\Big]\sigma_3\nonumber\\
&=&\hat{\gamma}^{\mu}(Q)-2g\sum_P\sum_{\sigma=\pm1}\sigma_3\hat{G}(P,\sigma\mu)\hat{\Gamma}^{\mu}(Q)\hat{G}(P+Q,\sigma\mu)\sigma_3.
\end{eqnarray}
To prove the gauge invariance of the above vertex $\hat{\Gamma}^{\mu}(Q)$, we substitute Eq.~(\ref{FD1}) into the GWI (\ref{FWI1}). After rearranging both sides, we only need to prove that
\begin{eqnarray}\label{dFD}
2i\Delta\sigma_2i\gamma_5=-2g\sum_P\sum_{\sigma=\pm1}\sigma_3\hat{G}(P,\sigma\mu)q_{\mu}\hat{\Gamma}^{\mu}(Q)\hat{G}(P+Q,\sigma\mu)\sigma_3.
\end{eqnarray}
Details of the proof of this equation is in Appendix.\ref{app:DE}. Therefore, the vertex given by the integral equation (\ref{FD1}) respects the GWI, or in other words, the theory is gauge invariant. Moreover, from the proof in Appendix.\ref{app:DE} we conclude that any truncation of the integral equation can not produce a gauge invariant vertex since terms of different orders of $g$ in Eq.~(\ref{tmp4}) cancel each other.

Interestingly, as pointed out also by Nambu \cite{Nambu60}, for the nonrelativistic BCS theory, the integral equation is not only consistent with the generalized Ward identity associated with the EM vertex but also consist with the GWIs associated with three other interaction vertices (as shown in Eq. (4.4) of ref.~\cite{Nambu60}). Moreover, the integral equation of the EM vertex is a vector equation while the GWI is a scalar equation, they have different degrees of freedom so there is no strict one-to-one correspondence between the solutions to the integral equation and the EM vertex respecting the GWI.

For the relativistic BCS theory, those conclusions should remain the same. Hence the integral equation (\ref{FD1}) is not equivalent to the GWI associated with the EM interaction. What is equivalent to the GWI is the contracted integral equation given by
\begin{eqnarray}\label{FD2}
q_{\mu}\hat{\Gamma}^{\mu}(Q)=q_{\mu}\hat{\gamma}^{\mu}(Q)-2g\sum_P\sum_{\sigma=\pm1}\sigma_3\hat{G}(P,\sigma\mu)q_{\mu}\hat{\Gamma}^{\mu}(Q)\hat{G}(P+Q,\sigma\mu)\sigma_3.
\end{eqnarray}
Since Eq.~(\ref{FD1}) satisfies the GWI, we can also derive the GWI from Eq.~(\ref{FD2}). Thus any vertex obeying the GWI (\ref{FWI1}) must satisfy Eq.~(\ref{FD2}) but not necessarily Eq.~(\ref{FD1}). Substituting Eq.~(\ref{FWI1}) into Eq.~(\ref{FD2}), the right-hand side becomes
\begin{eqnarray}
& &\slashed{q}\sigma_3-2g\sum_P\sum_{\sigma=\pm1}\sigma_3\hat{G}(P,\sigma\mu)\big(\sigma_3\hat{G}^{-1}(P+Q,\sigma\mu)-\hat{G}^{-1}(P,\sigma\mu)\sigma_3\big)\hat{G}(P+Q,\sigma\mu)\sigma_3\nonumber\\
&=&\slashed{q}\sigma_3-2g\sum_P\sum_{\sigma=\pm1}\big(\sigma_3\hat{G}(P,\sigma\mu)-\hat{G}(P,\sigma\mu)\sigma_3\big)\nonumber\\
&=&q_{\mu}\hat{\Gamma}^{\mu}(Q) \mbox{       (left-hand side)}.
\end{eqnarray}
Therefore it is the contracted integral equation (\ref{FD2}) that is equivalent to the GWI but not the integral equation (\ref{FD1}). A Comparison with the vertex $\hat{\Gamma}^{\mu}$ determined by the latter shows that the vertex $\hat{\Gamma}^{\prime\mu}$ determined by Eq.~(\ref{FD2}) may differ by a gauge transformation $\hat{\Gamma}^{\prime\mu}=\hat{\Gamma}^{\mu}+\hat{\chi}^{\mu}$, where $\hat{\chi}^{\mu}$ satisfies the Lorentz equation $q_{\mu}\hat{\chi}^{\mu}=0$. Such gauge transformations correspond to, for example, $\hat{\chi}_{\mu}=\partial_{\mu}\hat{f}$, where $\hat{f}$ is a matrix of harmonic functions in Nambu space.

In Section \ref{chap:5}, we derived the generalized vertex function $\hat{\mathbf{\Sigma}}$ in the generalized gauge space. In Nambu space we would like to investigate the vertex $\hat{\Gamma}^{\prime\mu}$ and show that the GWI (\ref{WI2}) will reduce to Eq.~(\ref{FWI1}). Eqs.~(\ref{D1D2}) can be written as
\begin{eqnarray}
\Delta_1=-\frac{\left|\begin{array}{cc}Q^{\mu}_{13} & Q_{12}\\ Q^{\mu}_{23} & \tilde{Q}_{22}\end{array}\right|}{\left|\begin{array}{cc}\tilde{Q}_{11} & Q_{12}\\ Q_{21} & \tilde{Q}_{22}\end{array}\right|}A_{\mu}=-\Pi^{\mu}_1A_{\mu}, , \mbox{  }\Delta_2=\frac{\left|\begin{array}{cc}Q^{\mu}_{13} & \tilde{Q}_{11}\\ Q^{\mu}_{23} & Q_{21}\end{array}\right|}{\left|\begin{array}{cc}\tilde{Q}_{11} & Q_{12}\\ Q_{21} & \tilde{Q}_{22}\end{array}\right|}A_{\mu}=\Pi^{\mu}_2A_{\mu},
\end{eqnarray}
where $\Pi^{\mu}_{1,2}$ satisfy $q_{\mu}\Pi^{\mu}_1=0$ and $q_{\mu}\Pi^{\mu}_2=-2i\Delta$ by noting Eq.(\ref{WI}). Therefore the gauge invariant vertex $\hat{\Gamma}^{\prime\mu}$ is given by
\begin{eqnarray}
\hat{\Gamma}^{\prime\mu}=\hat{\gamma}^{\mu}-\Pi^{\mu}_1\sigma_1i\gamma_5-\Pi^{\mu}_2\sigma_2i\gamma_5,
\end{eqnarray}
which obviously obeys the GWI
\begin{eqnarray}\label{GWI3}
q_{\mu}\hat{\Gamma}^{\prime\mu}=\slashed{q}\sigma_3-2\Delta\sigma_2\gamma_5=\sigma_3\hat{G}^{-1}(P+Q,\mu)-\hat{G}^{-1}(P,\mu)\sigma_3.
\end{eqnarray}
As we have discussed previously, since $\hat{\Gamma}^{\prime\mu}$ satisfies the GWI, it should obey Eq.~(\ref{FD2}). Hence it can differ from $\hat{\Gamma}^{\mu}$ given by Eq.~(\ref{FD1}) by a matrix function $\hat{\chi}^{\mu}$ at most. Moreover, the EM response kernel is now expressed as
\begin{eqnarray}\label{K2}
K^{\mu\nu}(Q)=\textrm{Tr}\sum_P\big(\hat{\Gamma}^{\prime\mu}\hat{G}(P+Q,\mu)\hat{\gamma}^{\nu}\hat{G}(P,\mu)\big),
\end{eqnarray}
where Eq.(\ref{dK}) has been used. Compare to the bare response kernel (\ref{K0}), the GWI (\ref{GWI3}) leads to the gauge invariance condition of the full response kernel $q_{\mu}K^{\mu\nu}(Q)=0$.

Although a solution to the integral equation gives a gauge invariant vertex, it is a great challenge to find a solution. We emphasize that the integral equation (\ref{FD1}) for the relativistic BCS theory should be implemented when one seeks gauge-invariant response functions. Previous work based on RPA approximations \cite{ReddyPRC04} implemented an iterative method without explicitly showing the complete integral equation for the relativistic BCS model. Further investigations of Nambu's integral-equation approach are needed for a better comparison between the results from the two approaches.

\section{Conclusion}\label{chap:conclusion}
The CFOP approach to the linear response functions of relativistic BCS superfluids restores the gauge invariance of the response functions of an external EM field. The manageable computability of this theory allows one to explore several interesting phenomena including collective modes, compressibility, and Meissner effect. Importantly, this approach leads to a consistent expression for the compressibility. When the pairing interaction is tunable, a BCS superfluid may exhibit a BCS - Bose-Einstein condensation (BEC) - relativistic BEC crossover \cite{OurNPA09}. Interesting issues may be raised in the linear response theory of a relativistic BCS superfluid in the crossover. Our CFOP approach could provide some useful tools for investigating those issues.

CCC acknowledges the support of the U. S. Department of Energy through the LANL/LDRD Program.

\appendix
\section{Spinor conventions} \label{app:spinor}
Here we use the Weyl or chiral representation of the $\gamma-$matrices,
\begin{equation}
\gamma^{0}=
\begin{pmatrix}\displaystyle
0 & I \\
I & 0
\end{pmatrix}
,\mbox{ } \gamma^{i}=-\gamma_{i}=
\begin{pmatrix}\displaystyle
0 & \sigma_{i} \\
-\sigma_{i} & 0
\end{pmatrix}
,\mbox{ } \gamma_{5}=
\begin{pmatrix}\displaystyle
-I & 0 \\
0 & I
\end{pmatrix}.
\end{equation}
The metric is chosen as $\eta_{\mu\nu}=\textrm{diag}(1,-1,-1,-1)$. The charge conjugation matrix $C$ is defined as $C=i\gamma^0\gamma^2$, which satisfies $C^{2}=-I$, $C^{\dagger}=C^{T}=-C$ and $[\gamma_{5},C]=0$. The $\gamma-$matrices satisfy $\gamma^{\mu*}=\gamma^{0}\gamma^{\mu T}\gamma^{0}$ and $C\gamma^{\mu T}C=\gamma^{\mu}$. The energy projectors satisfies
\begin{eqnarray}
\gamma^0\Lambda_{\pm}(\mathbf{p})&=&\Lambda_{\pm}(-\mathbf{p})\gamma^0,\nonumber\\
i\gamma_5\Lambda_{\pm}(\mathbf{p})&=&\Lambda_{\mp}(-\mathbf{p})i\gamma_5,\nonumber\\
\gamma^0i\gamma_5\Lambda_{\pm}(\mathbf{p})&=&\Lambda_{\mp}(\mathbf{p})\gamma^0i\gamma_5,
\end{eqnarray}
and
\begin{eqnarray}\bar{\sigma}_+\Lambda_-(\mathbf{p})+\bar{\sigma}_-\Lambda_-(\mathbf{p})+\bar{\sigma}_+\Lambda_+(\mathbf{p})+\bar{\sigma}_+\Lambda_-(\mathbf{p})=1.\end{eqnarray}

\section{The Fermion Propagator in Nambu Space}\label{appB}
From Eq.(\ref{eqn:FP}), the Fermion propagator in Nambu space is
\begin{eqnarray}
& &\hat{G}(P,\mu)=\big[\gamma^0(i\omega_n+\mu\sigma_3-\gamma^0\vec{\gamma}\cdot\mathbf{p}-m\gamma^0+\Delta i\gamma^0\gamma_5\sigma_1)\big]^{-1} \nonumber \\
&=&\big[(i\omega_n)^2-(k^2+m^2+\Delta^2+\mu^2)+2\mu\gamma^0(m+\vec{\gamma}\cdot\mathbf{p})\sigma_3\big]^{-1}\nonumber\\
& &\times(i\omega_n-\mu\sigma_3+\gamma^0\vec{\gamma}\cdot\mathbf{p}+m\gamma^0-\Delta i\gamma^0\gamma_5\sigma_1)\gamma^0.
\end{eqnarray}
This can be written in a more compact form as
\begin{eqnarray}\label{BG2}
& &\hat{G}(P,\mu)=\\
& &\frac{\big[(i\omega_n)^2-\epsilon^2_{\mathbf{p}}-\Delta^2-\mu^2-2\mu\gamma^0(m+\vec{\gamma}\cdot\mathbf{p})\sigma_3\big]\big[i\omega_n-\mu\sigma_3
+\gamma^0(m+\vec{\gamma}\cdot\mathbf{p})-\Delta i\gamma^0\gamma_5\sigma_1\big]\gamma^0}{\big((i\omega_n)^2-\epsilon^2_{\mathbf{p}}-\Delta^2-\mu^2\big)^2-4\mu^2\epsilon^2_{\mathbf{p}}} \nonumber
\end{eqnarray}
Note that $\gamma^0(m+\vec{\gamma}\cdot\mathbf{p})=\epsilon_{\mathbf{p}}\big(\Lambda_+(\mathbf{p})-\Lambda_-(\mathbf{p})\big)$ and $\sigma_3=\bar{\sigma}_+-\bar{\sigma}_-$. The first part of the numerator can be evaluated as
\begin{eqnarray}\label{app:B2}
& &(i\omega_n)^2-\epsilon^2_{\mathbf{p}}-\Delta^2-\mu^2-2\mu\gamma^0(m+\vec{\gamma}\cdot\mathbf{p})\sigma_3 \nonumber \\
&=&\big((i\omega_n)^2-E_{\mathbf{p}}^{+2}\big)\big(\bar{\sigma}_+\Lambda_+(\mathbf{p})+\bar{\sigma}_-\Lambda_-(\mathbf{p})\big)
+\big((i\omega_n)^2-E_{\mathbf{p}}^{-2}\big)\big(\bar{\sigma}_-\Lambda_+(\mathbf{p})+\bar{\sigma}_+\Lambda_-(\mathbf{p})\big).
\end{eqnarray}
The second part of the numerator can be evaluated as
\begin{eqnarray}
& &i\omega_n-\mu\sigma_3+\gamma^0(m+\vec{\gamma}\cdot\mathbf{p})-\Delta i\gamma^0\gamma_5\sigma_1 \nonumber \\
&=&(i\omega_n+\xi^-_{\mathbf{p}})\bar{\sigma}_+\Lambda_+(\mathbf{p})+(i\omega_n+\xi^+_{\mathbf{p}})\bar{\sigma}_-\Lambda_+(\mathbf{p})
+(i\omega_n-\xi^+_{\mathbf{p}})\bar{\sigma}_+\Lambda_-(\mathbf{p})+(i\omega_n-\xi^-_{\mathbf{p}})\bar{\sigma}_-\Lambda_-(\mathbf{p}). \nonumber
\end{eqnarray}
Note that $\xi^{\pm}_{\mathbf{p}}=(u^{\pm2}_{\mathbf{p}}-v^{\pm2}_{\mathbf{p}})E^{\pm}_{\mathbf{p}}$ and $u^{\pm2}_{\mathbf{p}}+v^{\pm2}_{\mathbf{p}}=1$. One has
\begin{eqnarray}\label{app:B3}
& &i\omega_n-\mu\sigma_3+\gamma^0(m+\vec{\gamma}\cdot\mathbf{p})-\Delta i\gamma^0\gamma_5\sigma_1   \nonumber \\
&=&\big[u^{-2}_{\mathbf{p}}(i\omega_n+E^-_{\mathbf{p}})+v^{-2}_{\mathbf{p}}(i\omega_n-E^-_{\mathbf{p}})\big]\bar{\sigma}_+\Lambda_+(\mathbf{p})\nonumber \\
&+&\big[u^{+2}_{\mathbf{p}}(i\omega_n+E^+_{\mathbf{p}})+v^{+2}_{\mathbf{p}}(i\omega_n-E^+_{\mathbf{p}})\big]\bar{\sigma}_-\Lambda_+(\mathbf{p})\nonumber \\
&+&\big[u^{+2}_{\mathbf{p}}(i\omega_n-E^+_{\mathbf{p}})+v^{+2}_{\mathbf{p}}(i\omega_n+E^+_{\mathbf{p}})\big]\bar{\sigma}_+\Lambda_-(\mathbf{p})\nonumber \\
&+&\big[u^{-2}_{\mathbf{p}}(i\omega_n-E^-_{\mathbf{p}})+v^{-2}_{\mathbf{p}}(i\omega_n+E^-_{\mathbf{p}})\big]\bar{\sigma}_-\Lambda_-(\mathbf{p}).
\end{eqnarray}
The denominator becomes
\begin{eqnarray}\label{app:B4}
\big((i\omega_n)^2-\epsilon^2_{\mathbf{p}}-\Delta^2-\mu^2\big)^2-4\mu^2\epsilon^2_{\mathbf{p}}=((i\omega_n)^2-E^{-2}_{\mathbf{p}})((i\omega_n)^2-E^{+2}_{\mathbf{p}}).
\end{eqnarray}
After substituting Eqs.~(\ref{app:B2}), (\ref{app:B3}) and (\ref{app:B4}) into the expression of $\hat{G}(P,\mu)$ and using $\bar{\sigma}_+\sigma_1=\sigma_+$ and $\bar{\sigma}_-\sigma_1=\sigma_-$, we have
\begin{eqnarray}\label{app:BG}
& &\hat{G}(P,\mu) \nonumber\\
&=&\Big[\big(\frac{u^{-2}_{\mathbf{p}}}{i\omega_n-E^-_{\mathbf{p}}}+\frac{v^{-2}_{\mathbf{p}}}{i\omega_n+E^-_{\mathbf{p}}}\big)\Lambda_+(\mathbf{p})
+\big(\frac{u^{+2}_{\mathbf{p}}}{i\omega_n+E^+_{\mathbf{p}}}+\frac{v^{+2}_{\mathbf{p}}}{i\omega_n-E^+_{\mathbf{p}}}\big)\Lambda_-(\mathbf{p})\Big]\gamma^0\bar{\sigma}_+ \nonumber\\
&+&\Big[\big(\frac{u^{+2}_{\mathbf{p}}}{i\omega_n-E^+_{\mathbf{p}}}+\frac{v^{+2}_{\mathbf{p}}}{i\omega_n+E^+_{\mathbf{p}}}\big)\Lambda_+(\mathbf{p})
+\big(\frac{u^{-2}_{\mathbf{p}}}{i\omega_n+E^-_{\mathbf{p}}}+\frac{v^{-2}_{\mathbf{p}}}{i\omega_n-E^-_{\mathbf{p}}}\big)\Lambda_-(\mathbf{p})\Big]\gamma^0\bar{\sigma}_- \nonumber\\
&+&\Big[\frac{\Lambda_+(\mathbf{p})\Delta }{(i\omega_n)^2-E^{-2}_{\mathbf{p}}}
+\frac{\Lambda_-(\mathbf{p})\Delta }{(i\omega_n)^2-E^{+2}_{\mathbf{p}}}\Big]i\gamma_5\sigma_+
+\Big[\frac{\Lambda_+(\mathbf{p})\Delta }{(i\omega_n)^2-E^{+2}_{\mathbf{p}}}
+\frac{\Lambda_-(\mathbf{p})\Delta }{(i\omega_n)^2-E^{-2}_{\mathbf{p}}}\Big]i\gamma_5\sigma_-.
\end{eqnarray}

\section{$\delta$-function operator and general properties of functions of $\epsilon-\hat{E}_{\mathbf{p}}$} \label{app:C}
We would like to evaluate an arbitrary function with the argument $\epsilon-\hat{E}_{\mathbf{p}}$, where $\hat{E}_{\mathbf{p}}=\gamma^0(\vec{\gamma}\cdot\mathbf{p}+m)-\mu\sigma_3-\Delta\gamma^0i\gamma_5\sigma_1$. Following the derivation of Eq.~(\ref{app:B3}) ($\mu\rightarrow-\mu\Rightarrow E^+_{\mathbf{p}}\leftrightarrow E^-_{\mathbf{p}}$ and $\gamma^0(\vec{\gamma}\cdot\mathbf{p}+m)\rightarrow-\gamma^0(\vec{\gamma}\cdot\mathbf{p}+m)\Rightarrow \Lambda_+(\mathbf{p})\leftrightarrow \Lambda_-(\mathbf{p})$), we have
\begin{eqnarray}\label{app:C1}
& &\epsilon-\hat{E}_{\mathbf{p}}  \nonumber \\
&=&(\epsilon+\xi^+_{\mathbf{p}})\bar{\sigma}_+\Lambda_-(\mathbf{p})+(\epsilon+\xi^-_{\mathbf{p}})\bar{\sigma}_-\Lambda_-(\mathbf{p})
+(\epsilon-\xi^-_{\mathbf{p}})\bar{\sigma}_+\Lambda_+(\mathbf{p})+(\epsilon-\xi^+_{\mathbf{p}})\bar{\sigma}_-\Lambda_+(\mathbf{p}) \nonumber \\
&+&\Delta\gamma^0i\gamma_5\sigma_1 \nonumber\\
&=&\big[u^{+2}_{\mathbf{p}}(\epsilon+E^+_{\mathbf{p}})+v^{+2}_{\mathbf{p}}(\epsilon-E^+_{\mathbf{p}})\big]\bar{\sigma}_+\Lambda_-(\mathbf{p})+\big[u^{-2}_{\mathbf{p}}(\epsilon+E^-_{\mathbf{p}})+v^{-2}_{\mathbf{p}}(\epsilon-E^-_{\mathbf{p}})\big]\bar{\sigma}_-\Lambda_-(\mathbf{p})\nonumber \\
&+&\big[u^{-2}_{\mathbf{p}}(\epsilon-E^-_{\mathbf{p}})+v^{-2}_{\mathbf{p}}(\epsilon+E^-_{\mathbf{p}})\big]\bar{\sigma}_+\Lambda_+(\mathbf{p})+\big[u^{+2}_{\mathbf{p}}(\epsilon-E^+_{\mathbf{p}})+v^{+2}_{\mathbf{p}}(\epsilon+E^+_{\mathbf{p}})\big]\bar{\sigma}_-\Lambda_+(\mathbf{p})\nonumber\\
&+&u^{+}_{\mathbf{p}}v^{+}_{\mathbf{p}}\big((\epsilon+ E^{+}_{\mathbf{p}})-(\epsilon-
E^{+}_{\mathbf{p}})\big)\big(\bar{\sigma}_+\Lambda_-(\mathbf{p})+\bar{\sigma}_-\Lambda_+(\mathbf{p})\big)\gamma^0i\gamma_5\sigma_1 \nonumber\\
&+&u^{-}_{\mathbf{p}}v^{-}_{\mathbf{p}}\big((\epsilon+ E^{-}_{\mathbf{p}})-(\epsilon-
E^{-}_{\mathbf{p}})\big)\big(\bar{\sigma}_+\Lambda_+(\mathbf{p})+\bar{\sigma}_-\Lambda_-(\mathbf{p})\big)\gamma^0i\gamma_5\sigma_1.
\end{eqnarray}
Explicitly, the four components of $\epsilon-\hat{E}_{\mathbf{p}}$ in Nambu space are
\begin{eqnarray}\label{eqn:C2}
\epsilon-\hat{E}_{\mathbf{p}}=\hat{u}^-_{\mathbf{p}}(\epsilon-E^-_{\mathbf{p}})+\hat{v}^-_{\mathbf{p}}(\epsilon+E^-_{\mathbf{p}})
+\hat{u}^+_{\mathbf{p}}(\epsilon+E^+_{\mathbf{p}})+\hat{v}^+_{\mathbf{p}}(\epsilon-E^+_{\mathbf{p}}),
\end{eqnarray}
where
\begin{eqnarray}
& &\hat{u}^-_{\mathbf{p}}=\left[ \begin{array}{ccc} u^{-2}_{\mathbf{p}}\Lambda_+(\mathbf{p}) & -u^{-}_{\mathbf{p}}v^{-}_{\mathbf{p}}\Lambda_+(\mathbf{p})\gamma^0i\gamma_5 \\ -u^{-}_{\mathbf{p}}v^{-}_{\mathbf{p}}\Lambda_-(\mathbf{p})\gamma^0i\gamma_5& v^{-2}_{\mathbf{p}}\Lambda_-(\mathbf{p}) \end{array}\right], \\
& &\hat{v}^-_{\mathbf{p}}=\left[ \begin{array}{ccc} v^{-2}_{\mathbf{p}}\Lambda_+(\mathbf{p}) & u^{-}_{\mathbf{p}}v^{-}_{\mathbf{p}}\Lambda_+(\mathbf{p})\gamma^0i\gamma_5 \\ u^{-}_{\mathbf{p}}v^{-}_{\mathbf{p}}\Lambda_-(\mathbf{p})\gamma^0i\gamma_5& u^{-2}_{\mathbf{p}}\Lambda_-(\mathbf{p}) \end{array}\right],\\
& &\hat{u}^+_{\mathbf{p}}=\left[ \begin{array}{ccc} u^{+2}_{\mathbf{p}}\Lambda_-(\mathbf{p}) & u^{+}_{\mathbf{p}}v^{+}_{\mathbf{p}}\Lambda_-(\mathbf{p})\gamma^0i\gamma_5 \\ u^{+}_{\mathbf{p}}v^{+}_{\mathbf{p}}\Lambda_+(\mathbf{p})\gamma^0i\gamma_5& v^{+2}_{\mathbf{p}}\Lambda_+(\mathbf{p}) \end{array}\right],\\
& &\hat{v}^+_{\mathbf{p}}=\left[ \begin{array}{ccc} v^{+2}_{\mathbf{p}}\Lambda_-(\mathbf{p}) & -u^{+}_{\mathbf{p}}v^{+}_{\mathbf{p}}\Lambda_-(\mathbf{p})\gamma^0i\gamma_5 \\ -u^{+}_{\mathbf{p}}v^{+}_{\mathbf{p}}\Lambda_+(\mathbf{p})\gamma^0i\gamma_5& u^{+2}_{\mathbf{p}}\Lambda_+(\mathbf{p}) \end{array}\right].
\end{eqnarray}
The four components of $\hat{E}_{\mathbf{p}}$ in Nambu space explicitly are $\hat{E}_{\mathbf{p}}=$
\begin{eqnarray}\label{eqn:E}
\left[ \begin{array}{ccc} (v^{+2}_{\mathbf{p}}-u^{+2}_{\mathbf{p}})E^+_{\mathbf{p}}\Lambda_-+(u^{-2}_{\mathbf{p}}-v^{-2}_{\mathbf{p}})E^-_{\mathbf{p}}\Lambda_+
& -2(E^+_{\mathbf{p}}u^{+}_{\mathbf{p}}v^{+}_{\mathbf{p}}\Lambda_-+E^-_{\mathbf{p}}u^{-}_{\mathbf{p}}v^{-}_{\mathbf{p}}\Lambda_+)\gamma^0i\gamma_5 \\-2(E^+_{\mathbf{p}}u^{+}_{\mathbf{p}}v^{+}_{\mathbf{p}}\Lambda_++E^-_{\mathbf{p}}u^{-}_{\mathbf{p}}v^{-}_{\mathbf{p}}\Lambda_-)\gamma^0i\gamma_5 & (u^{+2}_{\mathbf{p}}-v^{+2}_{\mathbf{p}})E^+_{\mathbf{p}}\Lambda_++(v^{-2}_{\mathbf{p}}-u^{-2}_{\mathbf{p}})E^-_{\mathbf{p}}\Lambda_-
\end{array}\right].
\end{eqnarray}
We have omitted the argument $\mathbf{p}$ of $\Lambda_{\pm}$. After comparing those expressions, one finds that
\begin{eqnarray}
& &\hat{u}^-_{\mathbf{p}}=\frac{(E^-_{\mathbf{p}}+\hat{E}_{\mathbf{p}})\hat{\Lambda}_+(\mathbf{p})}{2E^-_{\mathbf{p}}},\mbox{ }
\hat{v}^-_{\mathbf{p}}=\frac{(E^-_{\mathbf{p}}-\hat{E}_{\mathbf{p}})\hat{\Lambda}_+(\mathbf{p})}{2E^-_{\mathbf{p}}},\\
& &\hat{u}^+_{\mathbf{p}}=\frac{(E^+_{\mathbf{p}}-\hat{E}_{\mathbf{p}})\hat{\Lambda}_-(\mathbf{p})}{2E^+_{\mathbf{p}}},\mbox{ }
\hat{v}^+_{\mathbf{p}}=\frac{(E^+_{\mathbf{p}}+\hat{E}_{\mathbf{p}})\hat{\Lambda}_-(\mathbf{p})}{2E^+_{\mathbf{p}}}.
\end{eqnarray}
It can be verified that the operators $\hat{u}$ and $\hat{v}$ also satisfy the following properties
\begin{eqnarray}\label{app:Cuv1}
\hat{u}^{\pm2}_{\mathbf{p}}=\hat{u}^{\pm}_{\mathbf{p}},\mbox{   } \hat{v}^{\pm2}_{\mathbf{p}}=\hat{v}^{\pm}_{\mathbf{p}},
\end{eqnarray}
\begin{eqnarray}\label{app:Cuv2}
\hat{u}^{\pm}_{\mathbf{p}}\hat{v}^{\pm}_{\mathbf{p}}=\hat{v}^{\pm}_{\mathbf{p}}\hat{u}^{\pm}_{\mathbf{p}}
=\hat{u}^{\pm}_{\mathbf{p}}\hat{u}^{\mp}_{\mathbf{p}}=\hat{u}^{\pm}_{\mathbf{p}}\hat{v}^{\mp}_{\mathbf{p}}
=\hat{v}^{\pm}_{\mathbf{p}}\hat{u}^{\mp}_{\mathbf{p}}=\hat{v}^{\pm}_{\mathbf{p}}\hat{v}^{\mp}_{\mathbf{p}}=0.
\end{eqnarray}
After multiplying $\epsilon-\hat{E}_{\mathbf{p}}$ by itself, one gets
\begin{eqnarray}\label{eqn:C3}
(\epsilon-\hat{E}_{\mathbf{p}})^n=\hat{u}^-_{\mathbf{p}}(\epsilon-E^-_{\mathbf{p}})^n+\hat{v}^-_{\mathbf{p}}(\epsilon+E^-_{\mathbf{p}})^n
+\hat{u}^+_{\mathbf{p}}(\epsilon+E^+_{\mathbf{p}})^n+\hat{v}^+_{\mathbf{p}}(\epsilon-E^+_{\mathbf{p}})^n,
\end{eqnarray}
where $n$ is a positive integer. One can find that Eq.~(\ref{eqn:C3}) holds for $n=0$, too. Interestingly, from Eq.~(\ref{app:BG}) one can see that the case $n=-1$ is also valid by inspecting the expression of $G(P,\mu)\gamma^0$:
\begin{eqnarray}\label{eqn:C4}
(\epsilon-\hat{E}_{\mathbf{p}})^{-1}=\hat{u}^-_{\mathbf{p}}(\epsilon-E^-_{\mathbf{p}})^{-1}+\hat{v}^-_{\mathbf{p}}(\epsilon+E^-_{\mathbf{p}})^{-1}
+\hat{u}^+_{\mathbf{p}}(\epsilon+E^+_{\mathbf{p}})^{-1}+\hat{v}^+_{\mathbf{p}}(\epsilon-E^+_{\mathbf{p}})^{-1}.
\end{eqnarray}
Following the same argument,
\begin{eqnarray}\label{eqn:C5}
(\epsilon-\hat{E}_{\mathbf{p}})^{-n}=\hat{u}^-_{\mathbf{p}}(\epsilon-E^-_{\mathbf{p}})^{-n}+\hat{v}^-_{\mathbf{p}}(\epsilon+E^-_{\mathbf{p}})^{-n}
+\hat{u}^+_{\mathbf{p}}(\epsilon+E^+_{\mathbf{p}})^{-n}+\hat{v}^+_{\mathbf{p}}(\epsilon-E^+_{\mathbf{p}})^{-n}.
\end{eqnarray}
That means that Eq.~(\ref{eqn:C3}) holds for any integer $n$. For any function $F(\epsilon-\hat{E}_{\mathbf{p}})$
, we have the expansion
\begin{eqnarray} \label{app:CE5}
F(\epsilon-\hat{E}_{\mathbf{p}})=\hat{u}^{-}_{\mathbf{p}}F(\epsilon-E^-_{\mathbf{p}})+\hat{v}^{-}_{\mathbf{p}}F(\epsilon+E^-_{\mathbf{p}})
+\hat{u}^{+}_{\mathbf{p}}F(\epsilon+E^+_{\mathbf{p}})+\hat{v}^{+}_{\mathbf{p}}F(\epsilon-E^+_{\mathbf{p}}).
\end{eqnarray}
Thus,
\begin{eqnarray}\label{eqn:C6}
e^{-i(\epsilon-\hat{E}_{\mathbf{p}})t}=\hat{u}^-_{\mathbf{p}}e^{-i(\epsilon-E^-_{\mathbf{p}})t}+\hat{v}^-_{\mathbf{p}}e^{-i(\epsilon+E^-_{\mathbf{p}})t}
+\hat{u}^+_{\mathbf{p}}e^{-i(\epsilon+E^+_{\mathbf{p}})t}+\hat{v}^+_{\mathbf{p}}e^{-i(\epsilon-E^+_{\mathbf{p}})t}.
\end{eqnarray}
By Fourier transform we get
\begin{eqnarray}\label{eqn:C7}
\delta(\epsilon-\hat{E}_{\mathbf{p}})=\hat{u}^-_{\mathbf{p}}\delta(\epsilon-E^-_{\mathbf{p}})+\hat{v}^-_{\mathbf{p}}\delta(\epsilon+E^-_{\mathbf{p}})
+\hat{u}^+_{\mathbf{p}}\delta(\epsilon+E^+_{\mathbf{p}})+\hat{v}^+_{\mathbf{p}}\delta(\epsilon-E^+_{\mathbf{p}}).
\end{eqnarray}

\section{Integral equation of EM vertex and GWI}\label{app:DE}
Before proving that a vertex determined by the integral equation (\ref{FD1}) must obey the Ward identity (\ref{WI1}), or equivalently, verifying Eq.(\ref{dFD}), we prove the following identity.
\begin{eqnarray}\label{DE0}
2g\sum_P\sum_{\sigma=\pm1}\big(\sigma_3\hat{G}(P,\sigma\mu)-\hat{G}(P,\sigma\mu)\sigma_3\big)=-2i\Delta\sigma_2i\gamma_5=\hat{\Sigma}\sigma_3-\sigma_3\hat{\Sigma}.
\end{eqnarray}
The left-hand side of (\ref{DE0}) is
\begin{eqnarray}\label{DE1}
&=&4g\left(\begin{array}{cc} 0 & \sum_P\big(F(P,\mu)+F(P,-\mu)\big) \\ -\sum_P\big(F(P,\mu)+F(P,-\mu)\big) & 0 \end{array}\right)\nonumber\\
&=&4i\Delta g\left(\begin{array}{cc} 0 & \sum_{\mathbf{p}}\big(\frac{1-2f(E^{-}_{\mathbf{p}})}{2E^{-}_{\mathbf{p}}}
+\frac{1-2f(E^{+}_{\mathbf{p}})}{2E^{+}_{\mathbf{p}}}\big)\gamma_5 \\ -\sum_{\mathbf{p}}\big(\frac{1-2f(E^{-}_{\mathbf{p}})}{2E^{-}_{\mathbf{p}}}
+\frac{1-2f(E^{+}_{\mathbf{p}})}{2E^{+}_{\mathbf{p}}}\big)\gamma_5  & 0 \end{array}\right) \nonumber\\
&=&-2i\Delta\sigma_2i\gamma_5,
\end{eqnarray}
where the gap equation (\ref{EGAP}) has been used. From $\hat{\Sigma}=\hat{G}^{-1}_0(P,\mu)-\hat{G}^{-1}(P,\mu)$ one concludes that
\begin{eqnarray}\label{tmp2}
\hat{G}(P,\mu)\hat{G}^{-1}_0(P,\mu)=1+\hat{G}(P,\mu)\hat{\Sigma},\mbox{   }\hat{G}^{-1}_0(P,\mu)\hat{G}(P,\mu)=1+\hat{\Sigma}\hat{G}(P,\mu).
\end{eqnarray}
Now we turn to the proof of Eq.(\ref{dFD}) by considering the right-hand side. By substituting Eq.(\ref{FD1}) into the right-hand side and repeating the process, we get an iterative equation
\begin{eqnarray}\label{tmp4}
& &\mbox{RHS of Eq.~(\ref{dFD})}\nonumber\\
&=&-2g\sum_{P\sigma}\sigma_3\hat{G}_{\sigma}(P)q_{\mu}\gamma^{\mu}(Q)\hat{G}_{\sigma}(P+Q)\sigma_3\nonumber\\
& &+(2g)^2\sum_{P_1P_2\sigma}\sigma_3\hat{G}_{\sigma}(P_1)\sigma_3\hat{G}_{\sigma}(P_2)q_{\mu}\gamma^{\mu}(Q)\hat{G}_{\sigma}(P_2+Q)\sigma_3\hat{G}_{\sigma}(P_1+Q)\sigma_3+\cdots\nonumber\\
&=&\sum_{i=1}^{\infty}(-2g)^{i}\sum_{P_1\cdots P_i\sigma}\sigma_3\hat{G}_{\sigma}(P_1)\cdots\sigma_3\hat{G}_{\sigma}(P_i)q_{\mu}\gamma^{\mu}(Q)\hat{G}_{\sigma}(P_i+Q)\sigma_3\cdots\hat{G}_{\sigma}(P_1+Q)\sigma_3\nonumber\\
&=&\sum_{i=1}^{\infty}(-2g)^{i}\sum_{P_1\cdots P_i\sigma}\prod_{k=1}^{i}\big[\sigma_3\hat{G}_{\sigma}(P_k)\big]q_{\mu}\gamma^{\mu}(Q)\prod_{k=1}^{i}\big[\hat{G}_{\sigma}(P_{i+1-k})\sigma_3\big],
\end{eqnarray}
where we have defined $\hat{G}_{\sigma}(P)=\hat{G}(P,\sigma\mu)$ to shorten the expression. After inserting the Ward identity (\ref{FWI0}) for the bare EM vertex and using Eqs.(\ref{tmp2}), we get
\begin{eqnarray}
& &\mbox{RHS of Eq.(\ref{dFD})}\nonumber\\
&=&\sum_{i=1}^{\infty}(-2g)^{i}\sum_{P_1\cdots P_i\sigma}\sigma_3\hat{G}_{\sigma}(P_1)\cdots\sigma_3\hat{G}_{\sigma}(P_i)\sigma_3\hat{G}_{0\sigma}^{-1}(P_i+Q)\hat{G}_{\sigma}(P_i+Q)\sigma_3\cdots\hat{G}_{\sigma}(P_1+Q)\sigma_3\nonumber\\
&-&\sum_{i=1}^{\infty}(-2g)^{i}\sum_{P_1\cdots P_i\sigma}\sigma_3\hat{G}_{\sigma}(P_1)\cdots\sigma_3\hat{G}_{\sigma}(P_i)\hat{G}_{0\sigma}^{-1}(P_i)\sigma_3\hat{G}_{\sigma}(P_i+Q)\sigma_3\cdots\hat{G}_{\sigma}(P_1+Q)\sigma_3\nonumber\\
&=&-2g\sum_{P\sigma}\big(\sigma_3\hat{G}_{\sigma}(P)-\hat{G}_{\sigma}(P)\sigma_3\big)\nonumber\\
&+&\sum_{i=2}^{\infty}(-2g)^{i}\sum_{P_1\cdots P_i\sigma}\sigma_3\hat{G}_{\sigma}(P_1)\cdots\sigma_3\hat{G}_{\sigma}(P_i)\hat{G}_{\sigma}(P_{i-1}+Q)\sigma_3\cdots\hat{G}_{\sigma}(P_1+Q)\sigma_3\nonumber\\
&+&\sum_{i=1}^{\infty}(-2g)^{i}\sum_{P_1\cdots P_i\sigma}\sigma_3\hat{G}_{\sigma}(P_1)\cdots\sigma_3\hat{G}_{\sigma}(P_i)\sigma_3\hat{\Sigma}\hat{G}_{\sigma}(P_{i}+Q)\sigma_3\cdots\hat{G}_{\sigma}(P_1+Q)\sigma_3\nonumber\\
&-&\sum_{i=2}^{\infty}(-2g)^{i}\sum_{P_1\cdots P_i\sigma}\sigma_3\hat{G}_{\sigma}(P_1)\cdots\sigma_3\hat{G}_{\sigma}(P_{i-1})\hat{G}_{\sigma}(P_{i}+Q)\sigma_3\cdots\hat{G}_{\sigma}(P_1+Q)\sigma_3\nonumber\\
&-&\sum_{i=1}^{\infty}(-2g)^{i}\sum_{P_1\cdots P_i\sigma}\sigma_3\hat{G}_{\sigma}(P_1)\cdots\sigma_3\hat{G}_{\sigma}(P_i)\hat{\Sigma}\sigma_3\hat{G}_{\sigma}(P_{i}+Q)\sigma_3\cdots\hat{G}_{\sigma}(P_1+Q)\sigma_3.
\end{eqnarray}
By changing the dummy index $i\rightarrow i+1$ for the second and fourth summations, we get
\begin{eqnarray}
& &\mbox{RHS of Eq.(\ref{dFD})}\nonumber\\
&=&2i\Delta\sigma_2i\gamma_5\nonumber\\
&-&\sum_{i=1}^{\infty}(-2g)^{i}\sum_{P_1\cdots P_i\sigma}\prod_{k=1}^{i}\big[\sigma_3\hat{G}_{\sigma}(P_k)\big]2g\sum_{P_{i+1}}\big[\sigma_3\hat{G}_{\sigma}(P_{i+1})-\hat{G}_{\sigma}(P_{i+1})\sigma_3\big]\prod_{k=1}^{i}\big[\hat{G}_{\sigma}(P_{i+1-k})\sigma_3\big]\nonumber\\
&+&\sum_{i=1}^{\infty}(-2g)^{i}\sum_{P_1\cdots P_i\sigma}\prod_{k=1}^{i}\big[\sigma_3\hat{G}_{\sigma}(P_k)\big]\big[\sigma_3\hat{\Sigma}-\hat{\Sigma}\sigma_3\big]\prod_{k=1}^{i}\big[\hat{G}_{\sigma}(P_{i+1-k})\sigma_3\big]\nonumber\\
&=&2i\Delta\sigma_2i\gamma_5=\mbox{left-hand-side of Eq.(\ref{dFD})},
\end{eqnarray}
where Eq.(\ref{DE0}) has been used. Therefore we have proved that any vertex that satisfies the integral equation must also satisfy the Ward identity and hence must be gauge invariant.

\section{the odevities and symmetries of the response functions}\label{app:ob}
 Since the energy projectors satisfy $\gamma_5\Lambda_{\pm}(\mathbf{p})=\Lambda_{\mp}(-\mathbf{p})\gamma_5$, we have (See Eqs.(\ref{eqn:G}) and (\ref{eqn:F}))
\begin{eqnarray}\label{app:DG}
\gamma_5G(P,-\mu)&=&G(-P,\mu)\gamma_5,\nonumber\\
\gamma_5F(P,-\mu)&=&F(-P,\mu)\gamma_5.
\end{eqnarray}
The energy projectors also satisfy $\gamma_0\Lambda_{\pm}(\mathbf{p})=\Lambda_{\pm}(-\mathbf{p})\gamma_0$ so we can conclude that
\begin{eqnarray}\label{app:D3}
& &\gamma^0G(P,\mu)=G(\bar{P},\mu)\gamma^0,\nonumber \\
& &\gamma^0F(P,\mu)=-F(\bar{P},\mu)\gamma^0,
\end{eqnarray}
where $\bar{P}\equiv(i\omega_n,-\mathbf{p})$. The propagator in Nambu space (\ref{app:BG}) can be written as
\begin{eqnarray}
\hat{G}(P,\mu)=G(P,\mu)\bar{\sigma}_++G(P,-\mu)\bar{\sigma}_-+F(P,\mu)\sigma_++F(P,-\mu)\sigma_-.
\end{eqnarray}
Using Eqs.~(\ref{app:DG}), $\sigma_2\sigma_{\pm}=-\sigma_{\mp}\sigma_2$ and $\sigma_2\bar{\sigma}_{\pm}=\bar{\sigma}_{\mp}\sigma_2$
one can show that
\begin{eqnarray}\label{app:DF2}
\sigma_2\gamma_5\hat{G}(P,\mu)=\sigma_3\hat{G}(-P,\mu)\sigma_3\sigma_2\gamma_5.
\end{eqnarray}
Similarly, using Eqs.~(\ref{app:D3}), $\sigma_3\sigma_{\pm}=-\sigma_{\pm}\sigma_3$, and $\sigma_3\bar{\sigma}_{\pm}=\bar{\sigma}_{\pm}\sigma_3$
we have
\begin{eqnarray}\label{app:DF3}
\sigma_3\gamma^0\hat{G}(P,\mu)=\hat{G}(\bar{P},\mu)\sigma_3\gamma^0.
\end{eqnarray}
Thus we can analyze the odevity of the response functions of the four-momentum. Explicitly,
\begin{eqnarray}
Q_{ij}(i\Omega_{l},\mathbf{q})
&=&\sum_{P}\textrm{Tr}\big[(\sigma_2\gamma_5)^2\hat{\Sigma}_i\hat{G}(P+Q,\mu)\hat{\Sigma}_j\hat{G}(P,\mu)\big]\nonumber\\
&=&(-1)^{(2\delta^{1i}+2\delta^{1j}+\delta^{2i}+\delta^{2j})}\sum_{P}\textrm{Tr}\big[\hat{\Sigma}_i\hat{G}(-P-Q,\mu)\hat{\Sigma}_j\hat{G}(-P,\mu)\big]\nonumber\\
&=&(-1)^{(\delta^{2i}+\delta^{2j})}\sum_{P}\textrm{Tr}\big[\hat{\Sigma}_i\hat{G}(-P-Q,\mu)\hat{\Sigma}_j\hat{G}(-P,\mu)\big].
\end{eqnarray}
Changing variables by $P\rightarrow-P$, one gets
\begin{eqnarray}\label{D9}
Q_{ij}(i\Omega_{l},\mathbf{q})=(-1)^{(\delta^{2i}+\delta^{2j})}\sum_{P}\textrm{Tr}\big[\hat{\Sigma}_i\hat{G}(P-Q,\mu)\hat{\Sigma}_j\hat{G}(P,\mu)\big].
\end{eqnarray}
Therefore,
\begin{eqnarray}\label{O1}
Q_{ij}(i\Omega_{l},\mathbf{q})\equiv Q_{ij}(Q)=(-1)^{(\delta^{2i}+\delta^{2j})}Q_{ij}(-Q)\equiv (-1)^{(\delta^{2i}+\delta^{2j})}Q_{ij}(-i\Omega_{l},-\mathbf{q}).
\end{eqnarray}
Using the relation (\ref{D9}), one can show that
\begin{eqnarray}\label{D11}
Q_{ji}(i\Omega_{l},\mathbf{q})=(-1)^{(\delta^{2i}+\delta^{2j})}Q_{ij}(i\Omega_{l},\mathbf{q}).
\end{eqnarray}
Next we analyze the odevity of the response functions about the spatial components of the momentum. Using Eq.~(\ref{app:DF3}), for $i,j\neq 3$, one has
\begin{eqnarray}\label{O2}
Q_{ij}(i\Omega_{l},\mathbf{q})&=&\sum_{P}\textrm{Tr}\big[(\sigma_3\gamma^0)^2\hat{\Sigma}_i\hat{G}(P+Q,\mu)\hat{\Sigma}_j\hat{G}(P,\mu)\big]\nonumber\\
&=&\sum_{P}\textrm{Tr}\big[\hat{\Sigma}_i\hat{G}(\bar{P}+\bar{Q},\mu)\hat{\Sigma}_j\hat{G}(\bar{P},\mu)\big]\nonumber\\
&=&\sum_{P}\textrm{Tr}\big[\hat{\Sigma}_i\hat{G}(P+\bar{Q},\mu)\hat{\Sigma}_j\hat{G}(P,\mu)\big]\nonumber\\
&=&Q_{ij}(i\Omega_{l},-\mathbf{q}),
\end{eqnarray}
where in the third line we have changed variables by $P\rightarrow\bar{P}$. For $i=1,2;j=3$, with the help of Eq.~(\ref{D11}), we only need to consider the case with $i\leq j$. Thus
\begin{eqnarray}\label{O3}
Q^{\mu}_{i3}(i\Omega_{l},\mathbf{q})=(-1)^{1+\delta^{\mu0}}Q^{\mu}_{i3}(i\Omega_{l},-\mathbf{q}).
\end{eqnarray}
For $i=j=3$, we get
\begin{eqnarray}\label{O4}
Q^{\mu\nu}_{33}(i\Omega_{l},\mathbf{q})=(-1)^{\delta^{\mu0}+\delta^{\nu0}}Q^{\mu\nu}_{33}(i\Omega_{l},-\mathbf{q}).
\end{eqnarray}
From the odevities of the response functions about the four-momentum (see Eq.~(\ref{O1})) and the spatial momentum (see Eqs.~(\ref{O2}), (\ref{O3}), and (\ref{O4})), we derive the odevity of the response functions about the boson Matubara frequency as follows.
For $i=j=1,2$ we have
\begin{eqnarray}
Q_{ii}(i\Omega_{l},\mathbf{q})=Q_{ii}(-i\Omega_{l},\mathbf{q}).
\end{eqnarray}
For $i=j=3$  we have
\begin{eqnarray}
Q^{\mu\nu}_{33}(i\Omega_{l},\mathbf{q})=\left\{\begin{array}{cc} Q^{\mu\nu}_{33}(-i\Omega_{l},\mathbf{q}) & \textrm{if $\mu=\nu=0$ or $\mu=i$, $\nu=j$} \\ -Q^{\mu\nu}_{33}(-i\Omega_{l},\mathbf{q}) & \textrm{if $\mu=0$, $\nu=i$ or $\mu=i$, $\nu=0$}\end{array}\right.
\end{eqnarray}
For $i=1$, $j=2$ we have
\begin{eqnarray}
Q_{12}(i\Omega_{l},\mathbf{q})=-Q_{12}(-i\Omega_{l},\mathbf{q}).
\end{eqnarray}
For $i=1$, $j=3$ we have
\begin{eqnarray}
Q^{\mu}_{13}(i\Omega_{l},\mathbf{q})=\left\{\begin{array}{cc} Q^{\mu}_{13}(-i\Omega_{l},\mathbf{q}) & \textrm{if $\mu=0$}  \\ -Q^{\mu}_{13}(-i\Omega_{l},\mathbf{q}) & \textrm{if $\mu=i$ }\end{array}\right.
\end{eqnarray}
For $i=2$, $j=3$ we have
\begin{eqnarray}
Q^{\mu}_{23}(i\Omega_{l},\mathbf{q})=\left\{\begin{array}{cc} -Q^{\mu}_{23}(-i\Omega_{l},\mathbf{q}) & \textrm{if $\mu=0$}  \\ Q^{\mu}_{23}(-i\Omega_{l},\mathbf{q}) & \textrm{if $\mu=i$ }\end{array}\right.
\end{eqnarray}

\section{Expressions of the Coherence Coefficients}\label{app:CE}
For convenience, we introduce $k^{\mu}=(\epsilon_{\mathbf{p}},\mathbf{p})$ and $\bar{k}^{\mu}\equiv k_{\mu}=(\epsilon_{\mathbf{p}},-\mathbf{p})$ so the energy projectors can be rewritten as
\begin{eqnarray}\label{eqn:appD1}
& &\Lambda_{+}(\mathbf{p})=\frac{\slashed{k}+m}{2\epsilon_{\mathbf{p}}}\gamma^{0}=\gamma^{0}\frac{\slashed{\bar{k}}+m}{2\epsilon_{\mathbf{p}}}, \nonumber \\
& &\Lambda_{-}(\mathbf{p})=\frac{\slashed{\bar{k}}-m}{2\epsilon_{\mathbf{p}}}\gamma^{0}=\gamma^{0}\frac{\slashed{k}-m}{2\epsilon_{\mathbf{p}}},
\end{eqnarray}
which also satisfy
\begin{eqnarray}
& &\Lambda_{+}(\mathbf{p})\gamma_{5}=\gamma_{5}\frac{\slashed{k}-m}{2\epsilon_{\mathbf{p}}}\gamma^{0}=\gamma_{5}\gamma^{0}\frac{\slashed{\bar{k}}-m}{2\epsilon_{\mathbf{p}}}=\gamma_{5}\Lambda_{-}(-\mathbf{p}), \nonumber \\
& &\Lambda_{-}(\mathbf{p})\gamma_{5}=\gamma_{5}\frac{\slashed{\bar{k}}+m}{2\epsilon_{\mathbf{p}}}\gamma^{0}=\gamma_{5}\gamma^{0}\frac{\slashed{k}+m}{2\epsilon_{\mathbf{p}}}=\gamma_{5}\Lambda_{+}(-\mathbf{p})
\end{eqnarray}
We define $A(\mathbf{p},\mathbf{q})=\epsilon_{\mathbf{p}+\mathbf{q}}\epsilon_{\mathbf{p}}-\epsilon^2_{\mathbf{p}}-\mathbf{p}\cdot\mathbf{q}$ and $B(\mathbf{p},\mathbf{q})=\epsilon_{\mathbf{p}+\mathbf{q}}\epsilon_{\mathbf{p}}+\epsilon^2_{\mathbf{p}}+\mathbf{p}\cdot\mathbf{q}$, 
 which satisfy $A(\mathbf{p},\mathbf{q})=A(-\mathbf{p},-\mathbf{q})$ and $B(\mathbf{p},\mathbf{q})=B(-\mathbf{p},-\mathbf{q})$. By using the above relations and the the identity
\begin{equation}
\mbox{Tr}\big[(\slashed{k}^{\prime}+m)\gamma^{\mu}(\slashed{k}+m)\gamma^{\nu}\big]=4\big[k^{\prime\mu}k^{\nu}+k^{\prime\nu}k^{\mu}-g^{\mu\nu}(k\cdot k^{\prime}-m^{2})\big],
\end{equation}
we can show that
\begin{eqnarray}
& &(u^-u^-)_{11}=(v^-v^-)_{11}=\frac{1}{2}\Big(1-\frac{\xi^-_{\mathbf{p}+\mathbf{q}}\xi^-_{\mathbf{p}}-\Delta^2}{E^-_{\mathbf{p}+\mathbf{q}}E^-_{\mathbf{p}}}\Big)
\frac{B(\mathbf{p},\mathbf{q})}{\epsilon_{\mathbf{p}+\mathbf{q}}\epsilon_{\mathbf{p}}},\\
& &(u^-u^-)_{22}=(v^-v^-)_{22}=\frac{1}{2}\Big(1-\frac{\xi^-_{\mathbf{p}+\mathbf{q}}\xi^-_{\mathbf{p}}+\Delta^2}{E^-_{\mathbf{p}+\mathbf{q}}E^-_{\mathbf{p}}}\Big)
\frac{B(\mathbf{p},\mathbf{q})}{\epsilon_{\mathbf{p}+\mathbf{q}}\epsilon_{\mathbf{p}}}.
\end{eqnarray}
If $\mu=\nu=0$ or $\mu=i$, $\nu=j$, the $33$-component is given by
\begin{eqnarray}
& &(u^-u^-)^{\mu\nu}_{33}+(v^-v^-)^{\mu\nu}_{33} \nonumber \\
&=&\left\{ \begin{array}{ll}
\Big(1+\frac{\xi^-_{\mathbf{p}+\mathbf{q}}\xi^-_{\mathbf{p}}-\Delta^2}{E^-_{\mathbf{p}+\mathbf{q}}E^-_{\mathbf{p}}}\Big)\frac{B(\mathbf{p},\mathbf{q})}{\epsilon_{\mathbf{p}+\mathbf{q}}\epsilon_{\mathbf{p}}} & \textrm{if $\mu=\nu=0$} \\
\Big(1+\frac{\xi^-_{\mathbf{p}+\mathbf{q}}\xi^-_{\mathbf{p}}+\Delta^2}{E^-_{\mathbf{p}+\mathbf{q}}E^-_{\mathbf{p}}}\Big)\frac{(\mathbf{p}+\mathbf{q})^i\mathbf{p}^j+(\mathbf{p}+\mathbf{q})^j\mathbf{p}^i+\delta^{ij}A(\mathbf{p},\mathbf{q})}{\epsilon_{\mathbf{p}+\mathbf{q}}\epsilon_{\mathbf{p}}} & \textrm{if $\mu=i$ and $\nu=j$}
\end{array}\right.
\end{eqnarray}
If $\mu=0$, $\nu=i$ or $\mu=i$, $\nu=0$, the $33$-component is given by
\begin{eqnarray}
& &(u^-u^-)^{\mu\nu}_{33}-(v^-v^-)^{\mu\nu}_{33}=\frac{i}{2}\Big(\frac{\xi^-_{\mathbf{p}}}{E^-_{\mathbf{p}}}-\frac{\xi^-_{\mathbf{p}+\mathbf{q}}}{E^-_{\mathbf{p}+\mathbf{q}}}\Big)
\frac{B(\mathbf{p},\mathbf{q})}{\epsilon_{\mathbf{p}+\mathbf{q}}\epsilon_{\mathbf{p}}}.
\end{eqnarray}
\begin{eqnarray}
& &(u^-u^-)^{0}_{13}+(v^-v^-)^{0}_{13}=-\frac{\Delta(\xi^-_{\mathbf{p}}+\xi^-_{\mathbf{p}+\mathbf{q}})}{E^-_{\mathbf{p}}E^-_{\mathbf{p}+\mathbf{q}}}\frac{B(\mathbf{p},\mathbf{q})}{\epsilon_{\mathbf{p}+\mathbf{q}}\epsilon_{\mathbf{p}}},\\
& &(u^-u^-)^{i}_{13}-(v^-v^-)^{i}_{13}=-\Delta\Big(\frac{1}{E^-_{\mathbf{p}+\mathbf{q}}}+\frac{1}{E^-_{\mathbf{p}}}\Big)\frac{\epsilon_{\mathbf{p}+\mathbf{q}}\mathbf{p}^i+\epsilon_{\mathbf{p}}(\mathbf{p}+\mathbf{q})^i}{\epsilon_{\mathbf{p}+\mathbf{q}}\epsilon_{\mathbf{p}}}.
\end{eqnarray}
\begin{eqnarray}
& &(u^-u^-)^{0}_{23}-(v^-v^-)^{0}_{23}=i\Delta\Big(\frac{1}{E^-_{\mathbf{p}+\mathbf{q}}}-\frac{1}{E^-_{\mathbf{p}}}\Big)\frac{B(\mathbf{p},\mathbf{q})}{\epsilon_{\mathbf{p}+\mathbf{q}}\epsilon_{\mathbf{p}}},\\
& &(u^-u^-)^{i}_{23}+(v^-v^-)^{i}_{23}=\frac{i\Delta(\xi^-_{\mathbf{p}}-\xi^-_{\mathbf{p}+\mathbf{q}})}{E^-_{\mathbf{p}}E^-_{\mathbf{p}+\mathbf{q}}}\frac{(\mathbf{p}+\mathbf{q})^i\epsilon_{\mathbf{p}}+\mathbf{p}^i\epsilon_{\mathbf{p}+\mathbf{q}}}{\epsilon_{\mathbf{p}+\mathbf{q}}\epsilon_{\mathbf{p}}}.
\end{eqnarray}
\begin{eqnarray}
& &(u^-v^-)_{11}=(v^-u^-)_{11}=\frac{1}{2}\Big(1+\frac{\xi^-_{\mathbf{p}+\mathbf{q}}\xi^-_{\mathbf{p}}-\Delta^2}{E^-_{\mathbf{p}+\mathbf{q}}E^-_{\mathbf{p}}}\Big)
\frac{B(\mathbf{p},\mathbf{q})}{\epsilon_{\mathbf{p}+\mathbf{q}}\epsilon_{\mathbf{p}}}. \\
& &(u^-v^-)_{22}=(v^-u^-)_{22}=\frac{1}{2}\Big(1+\frac{\xi^-_{\mathbf{p}+\mathbf{q}}\xi^-_{\mathbf{p}}+\Delta^2}{E^-_{\mathbf{p}+\mathbf{q}}E^-_{\mathbf{p}}}\Big)
\frac{B(\mathbf{p},\mathbf{q})}{\epsilon_{\mathbf{p}+\mathbf{q}}\epsilon_{\mathbf{p}}}.
\end{eqnarray}
If $\mu=\nu=0$ or $\mu=i$, $\nu=j$, the 33-component is given by
\begin{eqnarray}
& &(u^-v^-)^{\mu\nu}_{33}+(v^-u^-)^{\mu\nu}_{33}\nonumber \\
&=&\left\{ \begin{array}{ll}
\Big(1-\frac{\xi^-_{\mathbf{p}+\mathbf{q}}\xi^-_{\mathbf{p}}-\Delta^2}{E^-_{\mathbf{p}+\mathbf{q}}E^-_{\mathbf{p}}}\Big)\frac{B(\mathbf{p},\mathbf{q})}{\epsilon_{\mathbf{p}+\mathbf{q}}\epsilon_{\mathbf{p}}} & \textrm{if $\mu=\nu=0$} \\
\Big(1-\frac{\xi^-_{\mathbf{p}+\mathbf{q}}\xi^-_{\mathbf{p}}+\Delta^2}{E^-_{\mathbf{p}+\mathbf{q}}E^-_{\mathbf{p}}}\Big)\frac{(\mathbf{p}+\mathbf{q})^i\mathbf{p}^j+(\mathbf{p}+\mathbf{q})^j\mathbf{p}^i+\delta^{ij}A(\mathbf{p},\mathbf{q})}{\epsilon_{\mathbf{p}+\mathbf{q}}\epsilon_{\mathbf{p}}} & \textrm{if $\mu=i$ and $\nu=j$}
\end{array}\right.
\end{eqnarray}
If $\mu=0$, $\nu=i$ or $\mu=i$, $\nu=0$, the 33-component is given by
\begin{eqnarray}
& &(u^-v^-)^{\mu\nu}_{33}-(v^-u^-)^{\mu\nu}_{33}=\Big(\frac{\xi^-_{\mathbf{p}+\mathbf{q}}}{E^-_{\mathbf{p}+\mathbf{q}}}-\frac{\xi^-_{\mathbf{p}}}{E^-_{\mathbf{p}}}\Big)\frac{\epsilon_{\mathbf{p}+\mathbf{q}}\mathbf{p}^i+\epsilon_{\mathbf{p}}(\mathbf{p}+\mathbf{q})^i}{\epsilon_{\mathbf{p}+\mathbf{q}}\epsilon_{\mathbf{p}}}.
\end{eqnarray}
\begin{eqnarray}
(u^-v^-)_{12}=-(v^-u^-)_{12}=-\frac{i}{2}\Big(\frac{\xi^-_{\mathbf{p}}}{E^-_{\mathbf{p}}}+\frac{\xi^-_{\mathbf{p}+\mathbf{q}}}{E^-_{\mathbf{p}+\mathbf{q}}}\Big)
\frac{B(\mathbf{p},\mathbf{q})}{\epsilon_{\mathbf{p}+\mathbf{q}}\epsilon_{\mathbf{p}}}.
\end{eqnarray}
\begin{eqnarray}
(u^-v^-)^{0}_{13}+(v^-u^-)^{0}_{13}&=&\frac{\Delta(\xi^-_{\mathbf{p}}+\xi^-_{\mathbf{p}+\mathbf{q}})}{E^-_{\mathbf{p}}E^-_{\mathbf{p}+\mathbf{q}}}\frac{B(\mathbf{p},\mathbf{q})}{\epsilon_{\mathbf{p}+\mathbf{q}}\epsilon_{\mathbf{p}}},\\
(u^-v^-)^{i}_{13}-(v^-u^-)^{i}_{13}&=&-\Delta\Big(\frac{1}{E^-_{\mathbf{p}+\mathbf{q}}}-\frac{1}{E^-_{\mathbf{p}}}\Big)\frac{\epsilon_{\mathbf{p}+\mathbf{q}}\mathbf{p}^i+\epsilon_{\mathbf{p}}(\mathbf{p}+\mathbf{q})^i}{\epsilon_{\mathbf{p}+\mathbf{q}}\epsilon_{\mathbf{p}}}.
\end{eqnarray}
\begin{eqnarray}
(u^-v^-)^{0}_{23}-(v^-u^-)^{0}_{23}&=&i\Delta\Big(\frac{1}{E^-_{\mathbf{p}+\mathbf{q}}}+\frac{1}{E^-_{\mathbf{p}}}\Big)\frac{B(\mathbf{p},\mathbf{q})}{\epsilon_{\mathbf{p}+\mathbf{q}}\epsilon_{\mathbf{p}}},\\
(u^-v^-)^{i}_{23}+(v^-u^-)^{i}_{23}&=&-\frac{i\Delta(\xi^-_{\mathbf{p}}-\xi^-_{\mathbf{p}+\mathbf{q}})}{E^-_{\mathbf{p}}E^-_{\mathbf{p}+\mathbf{q}}}\frac{(\mathbf{p}+\mathbf{q})^i\epsilon_{\mathbf{p}}+\mathbf{p}^i\epsilon_{\mathbf{p}+\mathbf{q}}}{\epsilon_{\mathbf{p}+\mathbf{q}}\epsilon_{\mathbf{p}}}.
\end{eqnarray}
\begin{eqnarray}
& &(u^+u^+)_{11}=(v^+v^+)_{11}=\frac{1}{2}\Big(1-\frac{\xi^+_{\mathbf{p}+\mathbf{q}}\xi^+_{\mathbf{p}}-\Delta^2}{E^+_{\mathbf{p}+\mathbf{q}}E^+_{\mathbf{p}}}\Big)
\frac{B(\mathbf{p},\mathbf{q})}{\epsilon_{\mathbf{p}+\mathbf{q}}\epsilon_{\mathbf{p}}}.\\
& &(u^+u^+)_{22}=(v^+v^+)_{22}=\frac{1}{2}\Big(1-\frac{\xi^+_{\mathbf{p}+\mathbf{q}}\xi^+_{\mathbf{p}}+\Delta^2}{E^+_{\mathbf{p}+\mathbf{q}}E^+_{\mathbf{p}}}\Big)
\frac{B(\mathbf{p},\mathbf{q})}{\epsilon_{\mathbf{p}+\mathbf{q}}\epsilon_{\mathbf{p}}}.
\end{eqnarray}
If $\mu=\nu=0$ or $\mu=i$, $\nu=j$, the 33-component is given by
\begin{eqnarray}
& &(u^+u^+)^{\mu\nu}_{33}+(v^+v^+)^{\mu\nu}_{33} \nonumber\\
&=&\left\{ \begin{array}{ll}
\Big(1+\frac{\xi^+_{\mathbf{p}+\mathbf{q}}\xi^+_{\mathbf{p}}-\Delta^2}{E^+_{\mathbf{p}+\mathbf{q}}E^+_{\mathbf{p}}}\Big)\frac{B(\mathbf{p},\mathbf{q})}{\epsilon_{\mathbf{p}+\mathbf{q}}\epsilon_{\mathbf{p}}} & \textrm{if $\mu=\nu=0$} \\
\Big(1+\frac{\xi^+_{\mathbf{p}+\mathbf{q}}\xi^+_{\mathbf{p}}+\Delta^2}{E^+_{\mathbf{p}+\mathbf{q}}E^+_{\mathbf{p}}}\Big)\frac{(\mathbf{p}+\mathbf{q})^i\mathbf{p}^j+(\mathbf{p}+\mathbf{q})^j\mathbf{p}^i+\delta^{ij}A(\mathbf{p},\mathbf{q})}{\epsilon_{\mathbf{p}+\mathbf{q}}\epsilon_{\mathbf{p}}} & \textrm{if $\mu=i$ and $\nu=j$}
\end{array}\right.
\end{eqnarray}
If $\mu=\nu=0$ or $\mu=i$, $\nu=j$, the 33-component is given by
\begin{eqnarray}
(u^+u^+)^{\mu\nu}_{33}-(v^+v^+)^{\mu\nu}_{33}=-\Big(\frac{\xi^+_{\mathbf{p}+\mathbf{q}}}{E^+_{\mathbf{p}+\mathbf{q}}}+\frac{\xi^+_{\mathbf{p}}}{E^+_{\mathbf{p}}}\Big)\frac{\epsilon_{\mathbf{p}+\mathbf{q}}\mathbf{p}^i+\epsilon_{\mathbf{p}}(\mathbf{p}+\mathbf{q})^i}{\epsilon_{\mathbf{p}+\mathbf{q}}\epsilon_{\mathbf{p}}}.
\end{eqnarray}
\begin{eqnarray}
(u^+u^+)_{12}=-(v^+v^+)_{12}=\frac{i}{2}\Big(\frac{\xi^+_{\mathbf{p}}}{E^+_{\mathbf{p}}}-\frac{\xi^+_{\mathbf{p}+\mathbf{q}}}{E^+_{\mathbf{p}+\mathbf{q}}}\Big)
\frac{B(\mathbf{p},\mathbf{q})}{\epsilon_{\mathbf{p}+\mathbf{q}}\epsilon_{\mathbf{p}}}.
\end{eqnarray}
\begin{eqnarray}
(u^+u^+)^{0}_{13}+(v^+v^+)^{0}_{13}&=&\frac{\Delta(\xi^+_{\mathbf{p}}+\xi^+_{\mathbf{p}+\mathbf{q}})}{E^+_{\mathbf{p}}E^+_{\mathbf{p}+\mathbf{q}}}\frac{B(\mathbf{p},\mathbf{q})}{\epsilon_{\mathbf{p}+\mathbf{q}}\epsilon_{\mathbf{p}}},\\
(u^+u^+)^{i}_{13}-(v^+v^+)^{i}_{13}&=&-\Delta\Big(\frac{1}{E^+_{\mathbf{p}+\mathbf{q}}}+\frac{1}{E^+_{\mathbf{p}}}\Big)\frac{\epsilon_{\mathbf{p}+\mathbf{q}}\mathbf{p}^i+\epsilon_{\mathbf{p}}(\mathbf{p}+\mathbf{q})^i}{\epsilon_{\mathbf{p}+\mathbf{q}}\epsilon_{\mathbf{p}}}.
\end{eqnarray}
\begin{eqnarray}
(u^+u^+)^{0}_{23}-(v^+v^+)^{0}_{23}&=&-i\Delta\Big(\frac{1}{E^+_{\mathbf{p}+\mathbf{q}}}-\frac{1}{E^+_{\mathbf{p}}}\Big)\frac{B(\mathbf{p},\mathbf{q})}{\epsilon_{\mathbf{p}+\mathbf{q}}\epsilon_{\mathbf{p}}},\\
(u^+u^+)^{i}_{23}+(v^+v^+)^{i}_{23}&=&\frac{i\Delta(\xi^+_{\mathbf{p}}-\xi^+_{\mathbf{p}+\mathbf{q}})}{E^+_{\mathbf{p}}E^+_{\mathbf{p}+\mathbf{q}}}\frac{(\mathbf{p}+\mathbf{q})^i\epsilon_{\mathbf{p}}+\mathbf{p}^i\epsilon_{\mathbf{p}+\mathbf{q}}}{\epsilon_{\mathbf{p}+\mathbf{q}}\epsilon_{\mathbf{p}}}.
\end{eqnarray}
\begin{eqnarray}
& &(u^+v^+)_{11}=(v^+u^+)_{11}=\frac{1}{2}\Big(1+\frac{\xi^+_{\mathbf{p}+\mathbf{q}}\xi^+_{\mathbf{p}}-\Delta^2}{E^+_{\mathbf{p}+\mathbf{q}}E^+_{\mathbf{p}}}\Big)
\frac{B(\mathbf{p},\mathbf{q})}{\epsilon_{\mathbf{p}+\mathbf{q}}\epsilon_{\mathbf{p}}}.\\
& &(u^+v^+)_{22}=(v^+u^+)_{22}=\frac{1}{2}\Big(1+\frac{\xi^+_{\mathbf{p}+\mathbf{q}}\xi^+_{\mathbf{p}}+\Delta^2}{E^+_{\mathbf{p}+\mathbf{q}}E^+_{\mathbf{p}}}\Big)
\frac{B(\mathbf{p},\mathbf{q})}{\epsilon_{\mathbf{p}+\mathbf{q}}\epsilon_{\mathbf{p}}}.
\end{eqnarray}
If $\mu=\nu=0$ or $\mu=i$, $\nu=j$, the 33-component is given by
\begin{eqnarray}
& &(u^+v^+)^{\mu\nu}_{33}+(v^+u^+)^{\mu\nu}_{33}\nonumber \\
&=&\left\{ \begin{array}{ll}
\Big(1-\frac{\xi^+_{\mathbf{p}+\mathbf{q}}\xi^+_{\mathbf{p}}-\Delta^2}{E^+_{\mathbf{p}+\mathbf{q}}E^+_{\mathbf{p}}}\Big)\frac{B(\mathbf{p},\mathbf{q})}{\epsilon_{\mathbf{p}+\mathbf{q}}\epsilon_{\mathbf{p}}} & \textrm{if $\mu=\nu=0$} \\
\Big(1-\frac{\xi^+_{\mathbf{p}+\mathbf{q}}\xi^+_{\mathbf{p}}+\Delta^2}{E^+_{\mathbf{p}+\mathbf{q}}E^+_{\mathbf{p}}}\Big)\frac{(\mathbf{p}+\mathbf{q})^i\mathbf{p}^j+(\mathbf{p}+\mathbf{q})^j\mathbf{p}^i+\delta^{ij}A(\mathbf{p},\mathbf{q})}{\epsilon_{\mathbf{p}+\mathbf{q}}\epsilon_{\mathbf{p}}} & \textrm{if $\mu=i$ and $\nu=j$}\end{array}\right.
\end{eqnarray}
If $\mu=0$, $\nu=i$ or $\mu=i$, $\nu=0$, the 33-component is given by
\begin{eqnarray}
& &(u^+v^+)^{\mu\nu}_{33}-(v^+u^+)^{\mu\nu}_{33}=-\Big(\frac{\xi^+_{\mathbf{p}+\mathbf{q}}}{E^+_{\mathbf{p}+\mathbf{q}}}-\frac{\xi^+_{\mathbf{p}}}{E^+_{\mathbf{p}}}\Big)\frac{\epsilon_{\mathbf{p}+\mathbf{q}}\mathbf{p}^i+\epsilon_{\mathbf{p}}(\mathbf{p}+\mathbf{q})^i}{\epsilon_{\mathbf{p}+\mathbf{q}}\epsilon_{\mathbf{p}}}.
\end{eqnarray}
\begin{eqnarray}
(u^+v^+)_{12}=-(v^+u^+)_{12}= -\frac{i}{2}\Big(\frac{\xi^+_{\mathbf{p}}}{E^+_{\mathbf{p}}}+\frac{\xi^+_{\mathbf{p}+\mathbf{q}}}{E^+_{\mathbf{p}+\mathbf{q}}}\Big)
\frac{B(\mathbf{p},\mathbf{q})}{\epsilon_{\mathbf{p}+\mathbf{q}}\epsilon_{\mathbf{p}}}.
\end{eqnarray}
\begin{eqnarray}
(u^+v^+)^{0}_{13}+(v^+u^+)^{0}_{13}&=&\frac{\Delta(\xi^+_{\mathbf{p}}+\xi^+_{\mathbf{p}+\mathbf{q}})}{E^+_{\mathbf{p}}E^+_{\mathbf{p}+\mathbf{q}}}\frac{B(\mathbf{p},\mathbf{q})}{\epsilon_{\mathbf{p}+\mathbf{q}}\epsilon_{\mathbf{p}}},\\
(u^+v^+)^{i}_{13}-(v^+u^+)^{i}_{13}&=&-\Delta\Big(\frac{1}{E^+_{\mathbf{p}+\mathbf{q}}}-\frac{1}{E^+_{\mathbf{p}}}\Big)\frac{\epsilon_{\mathbf{p}+\mathbf{q}}\mathbf{p}^i+\epsilon_{\mathbf{p}}(\mathbf{p}+\mathbf{q})^i}{\epsilon_{\mathbf{p}+\mathbf{q}}\epsilon_{\mathbf{p}}}.
\end{eqnarray}
\begin{eqnarray}
(u^+v^+)^{0}_{23}-(v^+u^+)^{0}_{23}&=&-i\Delta\Big(\frac{1}{E^+_{\mathbf{p}+\mathbf{q}}}+\frac{1}{E^+_{\mathbf{p}}}\Big)\frac{B(\mathbf{p},\mathbf{q})}{\epsilon_{\mathbf{p}+\mathbf{q}}\epsilon_{\mathbf{p}}},\\
(u^+v^+)^{i}_{23}+(v^+u^+)^{i}_{23}&=&-\frac{i\Delta(\xi^+_{\mathbf{p}}-\xi^+_{\mathbf{p}+\mathbf{q}})}{E^+_{\mathbf{p}}E^+_{\mathbf{p}+\mathbf{q}}}\frac{(\mathbf{p}+\mathbf{q})^i\epsilon_{\mathbf{p}}+\mathbf{p}^i\epsilon_{\mathbf{p}+\mathbf{q}}}{\epsilon_{\mathbf{p}+\mathbf{q}}\epsilon_{\mathbf{p}}}. \end{eqnarray}
Now we evaluate the ``mixed'' terms.
\begin{eqnarray}
& &(u^-u^+)_{11}=(v^-v^+)_{11}=\frac{1}{2}\Big(1-\frac{\xi^-_{\mathbf{p}+\mathbf{q}}\xi^+_{\mathbf{p}}+\Delta^2}{E^-_{\mathbf{p}+\mathbf{q}}E^+_{\mathbf{p}}}\Big)
\frac{A(\mathbf{p},\mathbf{q})}{\epsilon_{\mathbf{p}+\mathbf{q}}\epsilon_{\mathbf{p}}}.\\
& &(u^-u^+)_{22}=(v^-v^+)_{22}=\frac{1}{2}\Big(1-\frac{\xi^-_{\mathbf{p}+\mathbf{q}}\xi^+_{\mathbf{p}}-\Delta^2}{E^-_{\mathbf{p}+\mathbf{q}}E^+_{\mathbf{p}}}\Big)
\frac{A(\mathbf{p},\mathbf{q})}{\epsilon_{\mathbf{p}+\mathbf{q}}\epsilon_{\mathbf{p}}}.
\end{eqnarray}
If $\mu=\nu=0$ or $\mu=i$, $\nu=j$, the 33-component is given by
\begin{eqnarray}
& &(u^-u^+)^{\mu\nu}_{33}+(v^-v^+)^{\mu\nu}_{33}\nonumber\\
&=&\left\{ \begin{array}{ll}
\Big(1+\frac{\xi^-_{\mathbf{p}+\mathbf{q}}\xi^+_{\mathbf{p}}+\Delta^2}{E^-_{\mathbf{p}+\mathbf{q}}E^+_{\mathbf{p}}}\Big)\frac{A(\mathbf{p},\mathbf{q})}{\epsilon_{\mathbf{p}+\mathbf{q}}\epsilon_{\mathbf{p}}} & \textrm{if $\mu=\nu=0$} \\
-\Big(1+\frac{\xi^-_{\mathbf{p}+\mathbf{q}}\xi^+_{\mathbf{p}}-\Delta^2}{E^-_{\mathbf{p}+\mathbf{q}}E^+_{\mathbf{p}}}\Big)\frac{(\mathbf{p}+\mathbf{q})^i\mathbf{p}^j+(\mathbf{p}+\mathbf{q})^j\mathbf{p}^i-\delta^{ij}B(\mathbf{p},\mathbf{q})}{\epsilon_{\mathbf{p}+\mathbf{q}}\epsilon_{\mathbf{p}}} & \textrm{if $\mu=i$ and $\nu=j$}.
\end{array}\right.
\end{eqnarray}
If $\mu=0$, $\nu=i$ or $\mu=i$, $\nu=0$, the 33-component is given by
\begin{eqnarray}
& &(u^-u^+)^{\mu\nu}_{33}-(v^-v^+)^{\mu\nu}_{33}=\Big(\frac{\xi^-_{\mathbf{p}+\mathbf{q}}}{E^-_{\mathbf{p}+\mathbf{q}}}+\frac{\xi^+_{\mathbf{p}}}{E^+_{\mathbf{p}}}\Big)\frac{(\mathbf{p}+\mathbf{q})^i\epsilon_{\mathbf{p}}-\mathbf{p}^i\epsilon_{\mathbf{p}+\mathbf{q}}}{\epsilon_{\mathbf{p}+\mathbf{q}}\epsilon_{\mathbf{p}}}.
\end{eqnarray}
\begin{eqnarray}
(u^-u^+)_{12}=-(v^-v^+)_{12}= \frac{i}{2}\Big(\frac{\xi^+_{\mathbf{p}}}{E^+_{\mathbf{p}}}-\frac{\xi^-_{\mathbf{p}+\mathbf{q}}}{E^-_{\mathbf{p}+\mathbf{q}}}\Big)
\frac{A(\mathbf{p},\mathbf{q})}{\epsilon_{\mathbf{p}+\mathbf{q}}\epsilon_{\mathbf{p}}}.
\end{eqnarray}
\begin{eqnarray}
(u^-u^+)^{0}_{13}+(v^-v^+)^{0}_{13}&=&\frac{\Delta(\xi^-_{\mathbf{p}+\mathbf{q}}-\xi^+_{\mathbf{p}})}{E^-_{\mathbf{p}+\mathbf{q}}E^+_{\mathbf{p}}}\frac{A(\mathbf{p},\mathbf{q})}{\epsilon_{\mathbf{p}+\mathbf{q}}\epsilon_{\mathbf{p}}},\\
(u^-u^+)^{i}_{13}-(v^-v^+)^{i}_{13}&=&-\Delta\Big(\frac{1}{E^-_{\mathbf{p}+\mathbf{q}}}-\frac{1}{E^+_{\mathbf{p}}}\Big)\frac{(\mathbf{p}+\mathbf{q})^i\epsilon_{\mathbf{p}}-\mathbf{p}^i\epsilon_{\mathbf{p}+\mathbf{q}}}{\epsilon_{\mathbf{p}+\mathbf{q}}\epsilon_{\mathbf{p}}}.
\end{eqnarray}
\begin{eqnarray}
(u^-u^+)^{0}_{23}-(v^-v^+)^{0}_{23}&=&i\Delta\Big(\frac{1}{E^-_{\mathbf{p}+\mathbf{q}}}+\frac{1}{E^+_{\mathbf{p}}}\Big)\frac{A(\mathbf{p},\mathbf{q})}{\epsilon_{\mathbf{p}+\mathbf{q}}\epsilon_{\mathbf{p}}},\\
(u^-u^+)^{i}_{23}+(v^-v^+)^{i}_{23}&=&\frac{i\Delta(\xi^-_{\mathbf{p}+\mathbf{q}}+\xi^+_{\mathbf{p}})}{E^-_{\mathbf{p}+\mathbf{q}}E^+_{\mathbf{p}}}\frac{(\mathbf{p}+\mathbf{q})^i\epsilon_{\mathbf{p}}-\mathbf{p}^i\epsilon_{\mathbf{p}+\mathbf{q}}}{\epsilon_{\mathbf{p}+\mathbf{q}}\epsilon_{\mathbf{p}}}.
\end{eqnarray}
\begin{eqnarray}
& &(u^-v^+)_{11}=(v^-u^+)_{11}=\frac{1}{2}\Big(1+\frac{\xi^-_{\mathbf{p}+\mathbf{q}}\xi^+_{\mathbf{p}}+\Delta^2}{E^-_{\mathbf{p}+\mathbf{q}}E^+_{\mathbf{p}}}\Big)
\frac{A(\mathbf{p},\mathbf{q})}{\epsilon_{\mathbf{p}+\mathbf{q}}\epsilon_{\mathbf{p}}}.\\
& &(u^-v^+)_{22}=(v^-u^+)_{22}=\frac{1}{2}\Big(1+\frac{\xi^-_{\mathbf{p}+\mathbf{q}}\xi^+_{\mathbf{p}}-\Delta^2}{E^-_{\mathbf{p}+\mathbf{q}}E^+_{\mathbf{p}}}\Big)
\frac{A(\mathbf{p},\mathbf{q})}{\epsilon_{\mathbf{p}+\mathbf{q}}\epsilon_{\mathbf{p}}}.
\end{eqnarray}
If $\mu=\nu=0$ or $\mu=i$, $\nu=j$, the 33-component is given by
\begin{eqnarray}
& &(u^-v^+)^{\mu\nu}_{33}+(v^-u^+)^{\mu\nu}_{33}\nonumber\\
&=&\left\{ \begin{array}{ll}
\Big(1-\frac{\xi^-_{\mathbf{p}+\mathbf{q}}\xi^+_{\mathbf{p}}+\Delta^2}{E^-_{\mathbf{p}+\mathbf{q}}E^+_{\mathbf{p}}}\Big)\frac{A(\mathbf{p},\mathbf{q})}{\epsilon_{\mathbf{p}+\mathbf{q}}\epsilon_{\mathbf{p}}} & \textrm{if $\mu=\nu=0$} \\
-\Big(1-\frac{\xi^-_{\mathbf{p}+\mathbf{q}}\xi^+_{\mathbf{p}}-\Delta^2}{E^-_{\mathbf{p}+\mathbf{q}}E^+_{\mathbf{p}}}\Big)\frac{(\mathbf{p}+\mathbf{q})^i\mathbf{p}^j+(\mathbf{p}+\mathbf{q})^j\mathbf{p}^i-\delta^{ij}B(\mathbf{p},\mathbf{q})}{\epsilon_{\mathbf{p}+\mathbf{q}}\epsilon_{\mathbf{p}}} & \textrm{if $\mu=i$ and $\nu=j$}\end{array}\right.
\end{eqnarray}
If $\mu=0$, $\nu=i$ or $\mu=i$, $\nu=0$, the 33-component is given by
\begin{eqnarray}
& &(u^-v^+)^{\mu\nu}_{33}-(v^-u^+)^{\mu\nu}_{33}=\Big(\frac{\xi^-_{\mathbf{p}+\mathbf{q}}}{E^-_{\mathbf{p}+\mathbf{q}}}-\frac{\xi^+_{\mathbf{p}}}{E^+_{\mathbf{p}}}\Big)\frac{(\mathbf{p}+\mathbf{q})^i\epsilon_{\mathbf{p}}-\mathbf{p}^i\epsilon_{\mathbf{p}+\mathbf{q}}}{\epsilon_{\mathbf{p}+\mathbf{q}}\epsilon_{\mathbf{p}}}.
\end{eqnarray}
\begin{eqnarray}
(u^-v^+)_{12}=-(v^-u^+)_{12}= -\frac{i}{2}\Big(\frac{\xi^+_{\mathbf{p}}}{E^+_{\mathbf{p}}}+\frac{\xi^-_{\mathbf{p}+\mathbf{q}}}{E^-_{\mathbf{p}+\mathbf{q}}}\Big)
\frac{A(\mathbf{p},\mathbf{q})}{\epsilon_{\mathbf{p}+\mathbf{q}}\epsilon_{\mathbf{p}}}.
\end{eqnarray}
\begin{eqnarray}
(u^-v^+)^{0}_{13}+(v^-u^+)^{0}_{13}&=&\frac{\Delta(\xi^+_{\mathbf{p}}-\xi^-_{\mathbf{p}+\mathbf{q}})}{E^+_{\mathbf{p}}E^-_{\mathbf{p}+\mathbf{q}}}\frac{A(\mathbf{p},\mathbf{q})}{\epsilon_{\mathbf{p}+\mathbf{q}}\epsilon_{\mathbf{p}}},\\
(u^-v^+)^{i}_{13}-(v^-u^+)^{i}_{13}&=&-\Delta\Big(\frac{1}{E^-_{\mathbf{p}+\mathbf{q}}}+\frac{1}{E^+_{\mathbf{p}}}\Big)\frac{(\mathbf{p}+\mathbf{q})^i\epsilon_{\mathbf{p}}-\mathbf{p}^i\epsilon_{\mathbf{p}+\mathbf{q}}}{\epsilon_{\mathbf{p}+\mathbf{q}}\epsilon_{\mathbf{p}}}.
\end{eqnarray}
\begin{eqnarray}
(u^-v^+)^{0}_{23}-(v^-u^+)^{0}_{23}&=&i\Delta\Big(\frac{1}{E^-_{\mathbf{p}+\mathbf{q}}}-\frac{1}{E^+_{\mathbf{p}}}\Big)\frac{A(\mathbf{p},\mathbf{q})}{\epsilon_{\mathbf{p}+\mathbf{q}}\epsilon_{\mathbf{p}}},\\
(u^-v^+)^{i}_{23}+(v^-u^+)^{i}_{23}&=&-\frac{i\Delta(\xi^-_{\mathbf{p}+\mathbf{q}}+\xi^+_{\mathbf{p}})}{E^-_{\mathbf{p}+\mathbf{q}}E^+_{\mathbf{p}}}\frac{(\mathbf{p}+\mathbf{q})^i\epsilon_{\mathbf{p}}-\mathbf{p}^i\epsilon_{\mathbf{p}+\mathbf{q}}}{\epsilon_{\mathbf{p}+\mathbf{q}}\epsilon_{\mathbf{p}}}.
\end{eqnarray}
\begin{eqnarray}
& &(u^+u^-)_{11}=(v^+v^-)_{11}=\frac{1}{2}\Big(1-\frac{\xi^+_{\mathbf{p}+\mathbf{q}}\xi^-_{\mathbf{p}}+\Delta^2}{E^+_{\mathbf{p}+\mathbf{q}}E^-_{\mathbf{p}}}\Big)
\frac{A(\mathbf{p},\mathbf{q})}{\epsilon_{\mathbf{p}+\mathbf{q}}\epsilon_{\mathbf{p}}}.\\
& &(u^+u^-)_{22}=(v^+v^-)_{22}=\frac{1}{2}\Big(1-\frac{\xi^+_{\mathbf{p}+\mathbf{q}}\xi^-_{\mathbf{p}}-\Delta^2}{E^+_{\mathbf{p}+\mathbf{q}}E^-_{\mathbf{p}}}\Big)
\frac{A(\mathbf{p},\mathbf{q})}{\epsilon_{\mathbf{p}+\mathbf{q}}\epsilon_{\mathbf{p}}}.
\end{eqnarray}
If $\mu=\nu=0$ or $\mu=i$, $\nu=j$, the 33-component is given by
\begin{eqnarray}
& &(u^+u^-)^{\mu\nu}_{33}+(v^+v^-)^{\mu\nu}_{33}\nonumber\\
&=&\left\{ \begin{array}{ll}
\Big(1+\frac{\xi^+_{\mathbf{p}+\mathbf{q}}\xi^-_{\mathbf{p}}+\Delta^2}{E^+_{\mathbf{p}+\mathbf{q}}E^-_{\mathbf{p}}}\Big)\frac{A(\mathbf{p},\mathbf{q})}{\epsilon_{\mathbf{p}+\mathbf{q}}\epsilon_{\mathbf{p}}} & \textrm{if $\mu=\nu=0$} \\
-\Big(1+\frac{\xi^+_{\mathbf{p}+\mathbf{q}}\xi^-_{\mathbf{p}}-\Delta^2}{E^+_{\mathbf{p}+\mathbf{q}}E^-_{\mathbf{p}}}\Big)\frac{(\mathbf{p}+\mathbf{q})^i\mathbf{p}^j+(\mathbf{p}+\mathbf{q})^j\mathbf{p}^i-\delta^{ij}B(\mathbf{p},\mathbf{q})}{\epsilon_{\mathbf{p}+\mathbf{q}}\epsilon_{\mathbf{p}}} & \textrm{if $\mu=i$ and $\nu=j$}
\end{array}\right.
\end{eqnarray}
If $\mu=0$, $\nu=i$ or $\mu=i$, $\nu=0$, the 33-component is given by
\begin{eqnarray}
& &(u^+u^-)^{\mu\nu}_{33}-(v^+v^-)^{\mu\nu}_{33}=-\Big(\frac{\xi^+_{\mathbf{p}+\mathbf{q}}}{E^+_{\mathbf{p}+\mathbf{q}}}+\frac{\xi^-_{\mathbf{p}}}{E^-_{\mathbf{p}}}\Big)\frac{(\mathbf{p}+\mathbf{q})^i\epsilon_{\mathbf{p}}-\mathbf{p}^i\epsilon_{\mathbf{p}+\mathbf{q}}}{\epsilon_{\mathbf{p}+\mathbf{q}}\epsilon_{\mathbf{p}}}.
\end{eqnarray}
\begin{eqnarray}
(u^+u^-)_{12}=-(v^+v^-)_{12}= \frac{i}{2}\Big(\frac{\xi^-_{\mathbf{p}}}{E^-_{\mathbf{p}}}-\frac{\xi^+_{\mathbf{p}+\mathbf{q}}}{E^+_{\mathbf{p}+\mathbf{q}}}\Big)
\frac{A(\mathbf{p},\mathbf{q})}{\epsilon_{\mathbf{p}+\mathbf{q}}\epsilon_{\mathbf{p}}}.
\end{eqnarray}
\begin{eqnarray}
(u^+u^-)^{0}_{13}+(v^+v^-)^{0}_{13}&=&\frac{\Delta(\xi^-_{\mathbf{p}}-\xi^+_{\mathbf{p}+\mathbf{q}})}{E^+_{\mathbf{p}+\mathbf{q}}E^-_{\mathbf{p}}}\frac{A(\mathbf{p},\mathbf{q})}{\epsilon_{\mathbf{p}+\mathbf{q}}\epsilon_{\mathbf{p}}},\\
(u^+u^-)^{i}_{13}-(v^+v^-)^{i}_{13}&=&-\Delta\Big(\frac{1}{E^+_{\mathbf{p}+\mathbf{q}}}-\frac{1}{E^-_{\mathbf{p}}}\Big)\frac{(\mathbf{p}+\mathbf{q})^i\epsilon_{\mathbf{p}}-\mathbf{p}^i\epsilon_{\mathbf{p}+\mathbf{q}}}{\epsilon_{\mathbf{p}+\mathbf{q}}\epsilon_{\mathbf{p}}}.
\end{eqnarray}
\begin{eqnarray}
(u^+u^-)^{0}_{23}-(v^+v^-)^{0}_{23}&=&-i\Delta\Big(\frac{1}{E^+_{\mathbf{p}+\mathbf{q}}}+\frac{1}{E^-_{\mathbf{p}}}\Big)\frac{A(\mathbf{p},\mathbf{q})}{\epsilon_{\mathbf{p}+\mathbf{q}}\epsilon_{\mathbf{p}}},\\
(u^+u^-)^{i}_{23}+(v^+v^-)^{i}_{23}&=&\frac{i\Delta(\xi^+_{\mathbf{p}+\mathbf{q}}+\xi^-_{\mathbf{p}})}{E^+_{\mathbf{p}+\mathbf{q}}E^-_{\mathbf{p}}}\frac{(\mathbf{p}+\mathbf{q})^i\epsilon_{\mathbf{p}}-\mathbf{p}^i\epsilon_{\mathbf{p}+\mathbf{q}}}{\epsilon_{\mathbf{p}+\mathbf{q}}\epsilon_{\mathbf{p}}}.
\end{eqnarray}
\begin{eqnarray}
& &(u^+v^-)_{11}=(v^+u^-)_{11}=\frac{1}{2}\Big(1+\frac{\xi^+_{\mathbf{p}+\mathbf{q}}\xi^-_{\mathbf{p}}+\Delta^2}{E^+_{\mathbf{p}+\mathbf{q}}E^-_{\mathbf{p}}}\Big)
\frac{A(\mathbf{p},\mathbf{q})}{\epsilon_{\mathbf{p}+\mathbf{q}}\epsilon_{\mathbf{p}}}.\\
& &(u^+v^-)_{22}=(v^+u^-)_{22}=\frac{1}{2}\Big(1+\frac{\xi^+_{\mathbf{p}+\mathbf{q}}\xi^-_{\mathbf{p}}-\Delta^2}{E^+_{\mathbf{p}+\mathbf{q}}E^-_{\mathbf{p}}}\Big)
\frac{A(\mathbf{p},\mathbf{q})}{\epsilon_{\mathbf{p}+\mathbf{q}}\epsilon_{\mathbf{p}}}.
\end{eqnarray}
If $\mu=\nu=0$ or $\mu=i$, $\nu=j$, the 33-component is given by
\begin{eqnarray}
& &(u^+v^-)^{\mu\nu}_{33}+(v^+u^-)^{\mu\nu}_{33}\nonumber\\
&=&\left\{ \begin{array}{ll}
\Big(1-\frac{\xi^+_{\mathbf{p}+\mathbf{q}}\xi^-_{\mathbf{p}}+\Delta^2}{E^+_{\mathbf{p}+\mathbf{q}}E^-_{\mathbf{p}}}\Big)\frac{A(\mathbf{p},\mathbf{q})}{\epsilon_{\mathbf{p}+\mathbf{q}}\epsilon_{\mathbf{p}}} & \textrm{if $\mu=\nu=0$} \\
-\Big(1-\frac{\xi^+_{\mathbf{p}+\mathbf{q}}\xi^-_{\mathbf{p}}-\Delta^2}{E^+_{\mathbf{p}+\mathbf{q}}E^-_{\mathbf{p}}}\Big)\frac{(\mathbf{p}+\mathbf{q})^i\mathbf{p}^j+(\mathbf{p}+\mathbf{q})^j\mathbf{p}^i-\delta^{ij}B(\mathbf{p},\mathbf{q})}{\epsilon_{\mathbf{p}+\mathbf{q}}\epsilon_{\mathbf{p}}} & \textrm{if $\mu=i$ and $\nu=j$}
\end{array}\right.
\end{eqnarray}
If $\mu=0$, $\nu=i$ or $\mu=i$, $\nu=0$, the 33-component is given by
\begin{eqnarray}
& &(u^+v^-)^{\mu\nu}_{33}-(v^+u^-)^{\mu\nu}_{33}=-\Big(\frac{\xi^+_{\mathbf{p}+\mathbf{q}}}{E^+_{\mathbf{p}+\mathbf{q}}}-\frac{\xi^-_{\mathbf{p}}}{E^-_{\mathbf{p}}}\Big)\frac{(\mathbf{p}+\mathbf{q})^i\epsilon_{\mathbf{p}}-\mathbf{p}^i\epsilon_{\mathbf{p}+\mathbf{q}}}{\epsilon_{\mathbf{p}+\mathbf{q}}\epsilon_{\mathbf{p}}}.
\end{eqnarray}
\begin{eqnarray}
(u^+v^-)_{12}=-(v^+u^-)_{12}= -\frac{i}{2}\Big(\frac{\xi^-_{\mathbf{p}}}{E^-_{\mathbf{p}}}+\frac{\xi^+_{\mathbf{p}+\mathbf{q}}}{E^+_{\mathbf{p}+\mathbf{q}}}\Big)
\frac{A(\mathbf{p},\mathbf{q})}{\epsilon_{\mathbf{p}+\mathbf{q}}\epsilon_{\mathbf{p}}}.
\end{eqnarray}
\begin{eqnarray}
(u^+v^-)^{0}_{13}+(v^+u^-)^{0}_{13}&=&\frac{\Delta(\xi^+_{\mathbf{p}+\mathbf{q}}-\xi^-_{\mathbf{p}})}{E^+_{\mathbf{p}+\mathbf{q}}E^-_{\mathbf{p}}}\frac{A(\mathbf{p},\mathbf{q})}{\epsilon_{\mathbf{p}+\mathbf{q}}\epsilon_{\mathbf{p}}},\\
(u^+v^-)^{i}_{13}-(v^+u^-)^{i}_{13}&=&-\Delta\Big(\frac{1}{E^+_{\mathbf{p}+\mathbf{q}}}+\frac{1}{E^-_{\mathbf{p}}}\Big)\frac{(\mathbf{p}+\mathbf{q})^i\epsilon_{\mathbf{p}}-\mathbf{p}^i\epsilon_{\mathbf{p}+\mathbf{q}}}{\epsilon_{\mathbf{p}+\mathbf{q}}\epsilon_{\mathbf{p}}}.
\end{eqnarray}
\begin{eqnarray}
(u^+v^-)^{0}_{23}-(v^+u^-)^{0}_{23}&=&-i\Delta\Big(\frac{1}{E^+_{\mathbf{p}+\mathbf{q}}}-\frac{1}{E^-_{\mathbf{p}}}\Big)\frac{A(\mathbf{p},\mathbf{q})}{\epsilon_{\mathbf{p}+\mathbf{q}}\epsilon_{\mathbf{p}}},\\
(u^+v^-)^{i}_{23}+(v^+u^-)^{i}_{23}&=&-\frac{i\Delta(\xi^+_{\mathbf{p}+\mathbf{q}}+\xi^-_{\mathbf{p}})}{E^+_{\mathbf{p}+\mathbf{q}}E^-_{\mathbf{p}}}\frac{(\mathbf{p}+\mathbf{q})^i\epsilon_{\mathbf{p}}-\mathbf{p}^i\epsilon_{\mathbf{p}+\mathbf{q}}}{\epsilon_{\mathbf{p}+\mathbf{q}}\epsilon_{\mathbf{p}}}. \end{eqnarray}

\vspace*{-1ex}

\bibliographystyle{apsrev}

\end{document}